# Revisiting multiple trapping and release electronic transport in amorphous semiconductors exemplified by *a*-Si:H

Yuezhou Luo* and Andrew John Flewitt

*Electrical Engineering Division, Department of Engineering, University of Cambridge, Cambridge CB3 0FA, United Kingdom*

Multiple trapping and release (MTR) is a typical electronic transport mechanism associated with localized states in amorphous and other disordered semiconductors. For instance, it dominates the room-temperature charge transport in hydrogenated amorphous silicon (*a*-Si:H). However, till now analysis of MTR transport has been built on an "abrupt" mobility edge model. Using electron transport as an example, the abrupt mobility edge model assumes that: (i) states above the conduction band (CB) mobility edge ($E_C$) are extended and any of them is omnipresent in space, whereas states below $E_C$ are localized and they exist in the energy-space diagram as pointlike sites; (ii) all states are evenly distributed in space, and the local density of states (DOS) distribution is spatially invariant. For a steady-state conduction [such as DC conduction and non-dispersive time-of-flight (TOF) conduction], this abrupt mobility edge model quantifies the average electron mobility of a disordered semiconductor as a fraction of the extended-state electron mobility ($\mu_C$); the fraction is associated with the density ratio between extended-state electrons and localized-state electrons. Not doubted before, based on this model, the two key quantities, $E_C$ and $\mu_C$, were explicitly determined through conductivity activation energy measurement, DOS data or models, and TOF experiment. The widely accepted view was that the determined $E_C$ exactly demarcates extended states and localized states, and that the determined $\mu_C$ represents the mobility of free electrons in the disordered semiconductor.

The prequel to this paper [Y. Luo and A. J. Flewitt, Phys. Rev. B **109**, 104203 (2024)] demonstrates that neither of the aforementioned assumptions is valid. Hence, this paper reinvestigates MTR transport. Through a probabilistic analysis of the microscopic charge transport details, this paper rigorously demonstrates that, first, the experimentally estimated $E_C$ is an effective mobility edge which is different from the critical energy that demarcates extended states and localized states in a disordered semiconductor. Second, the experimentally determined $\mu_C$ turns out to be higher than the actual mobility of free electrons in the material. The *a*-Si:H discussed in the prequel, being an intensively studied sample in the past, is used as an example to concretize the analysis. The effective mobility edge of this *a*-Si:H is 0.074 eV higher than the critical energy. Further, the extended-state mobility of this material is estimated to be ~ 46 cm$^2$/(V s) through fitting to temperature-dependent TOF data of this sample; this is indeed higher than the mobility of free electrons in *a*-Si:H which is only ~ 5 cm$^2$/(V s) as approximated by the upper limit of Brownian motion.

## I. INTRODUCTION

Understanding charge transport mechanisms in amorphous semiconductors has been a matter of significant interest since the last century as a result of the technological importance of this class of materials. Trap controlled transport, also known as multiple trapping and release (MTR) transport, has proved crucial in amorphous semiconductors [1-8]. In this mechanism, charged carriers interact with the localized states and extended states of an amorphous semiconductor through trapping and thermal release events, thereby migrating and carrying net current through the material. In 1970, Le Comber and Spear [4] analyzed the electron drift mobility data from time-of-flight (TOF) measurements and concluded that near and above room temperature, the electronic transport in a hydrogenated amorphous silicon (*a*-Si:H) sample is predominantly trap controlled, whereas hopping is much less significant.

As shown in Fig. 1, a stereotypical model was widely adopted to describe the process of MTR electron transport, where an abrupt mobility edge ($E_C$) exists in the conduction band (CB). Electrons are in extended states above this edge and exhibit an extended-state mobility $\mu_C$, whereas electrons are in localized states below this edge and possess zero mobility. During a TOF experiment, a photogenerated electron carrier may be trapped from $E_C$ into a localized state at the $i^{\text{th}}$ level below $E_C$. Through energy exchange, after a certain duration this electron may be thermally released to a high-energy extended state at the $j^{\text{th}}$ level above $E_C$ and then rapidly relaxes to and drift at $E_C$. Such trapping and release processes continues till the photogenerated electron completely transits through the material. Under this framework, the electron drift mobility of *a*-Si:H measured from a TOF experiment near room temperature is [9]

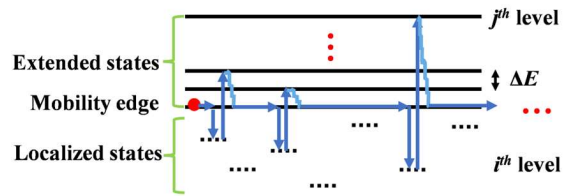

FIG. 1. Illustration of multiple trapping and release (MTR) electronic transport based on the model of abrupt mobility edge. The red sphere denotes a representative electron carrier. Dotted black line segments denote localized states below the conduction band (CB) mobility edge ($E_C$), while the solid black lines represent extended states. Energy levels are quasicontinuously discretized with an energy interval of $\Delta E$. The zig-zagged light blue lines denotes the rapid relaxation processes. $i$ and $j$ are the indices for localized-state and extended-state energy levels, respectively.

$$\mu_d = \mu_C \frac{t_{ext}}{t_{ext} + t_{loc}}, \qquad (1)$$

where $t_{ext}$ denotes the cumulative time that the representative electron spends in extended states, and $t_{loc}$ denotes the cumulative time that the electron spends in localized states.

MTR transport in disordered semiconductors during a TOF experiment is generally classified into dispersive transport and non-dispersive transport. Non-dispersive transport means that the measured drift mobility of carriers does not vary with applied electric field or sample thickness. This, however, does not hold true for dispersive transport where a process called progressive thermalization ([10,11]) is significant relative to the TOF measurement duration. During the progressive thermalization, a demarcation energy [$E_d(\tau)$] progressively sinks down from $E_C$ towards the dark Fermi level $E_F$, which follows [10,11]

* YL778@cam.ac.uk

$$E_d(\tau) = E_C - kT \ln(\nu\tau), \quad (2)$$

for $\tau \geq \nu^{-1}$, where $\tau$ denotes time and $\nu$ is an attempt-to-escape frequency of the order of $10^{12} \sim 10^{13}$ s$^{-1}$ [11]. $k$ is the Boltzmann constant, and $T$ denotes temperature. The dynamic sinking of demarcation energy varies the distribution of the photogenerated electrons during the TOF transit, because above the demarcation energy the occupation probability of these electrons follows a Boltzmann distribution of the form [10-12]

$$F(E,\tau) = F_s(\tau) \exp\left[\frac{E_d(\tau) - E}{kT}\right], E > E_d(\tau), \quad (3)$$

whereas below the demarcation energy the photogenerated electrons remain "frozen" in the form they were initially trapped to these localized states [10]. Based on the understanding of Ref. [10], depending on the electron capture cross section of these localized states, the occupation probability is quantified as

$$F(E,\tau) = F_s(\tau) \frac{\sigma(E_C, E)}{\sigma[E_C, E_d(\tau)]}, E < E_d(\tau), \quad (4)$$

where the capture cross section $\sigma(E_C, E)$ quantifies the ease of capturing a free electron from $E_C$ to an energy level $E$. If the capture cross section is constant, the distribution of photogenerated electrons below the demarcation energy will parallel the density of states (DOS) distribution [10]. In both Eq. (3) and Eq. (4) the scaling factor $F_s(\tau)$ also evolves with time so that the total number of photogenerated electrons is conserved.

Until either of the following cases occurs, the above progressive thermalization process will significantly vary the energetic distribution of photogenerated electrons:

(1) After a duration of $t_0$, the demarcation energy sinks to an energy $E_b$, below which the DOS is trivial and keeps decreasing rapidly with a decreasing energy [10].

(2) After a duration of $t_d$, the demarcation energy sinks to an energy $E_D$, below which $\sigma(E_C, E)$ drops abruptly [13-15].

After Case (1) is reached, the continuous sinking of demarcation energy thermalizes only a very trivial number of previously frozen electrons; meanwhile the number of localized states between $E_d(\tau)$ and $E_b$ is too small to alter the electron distribution above $E_b$. As a result, the energetic distribution of unfrozen photogenerated electrons starts to appear stable. For Case (2), since the capture cross section below $E_D$ is considerably small, according to Eq. (4), almost none of the photogenerated electrons at the start of the TOF experiment would ever fall into the states below $E_D$ throughout the TOF duration. Therefore, after the demarcation energy sinks to $E_D$, all photogenerated electrons will have been thermalized and their occupation follows a stable Boltzmann distribution.

For either case, if $t_0$ or $t_d$ is comparable or longer than the total transit time of the group of photogenerated electrons, the TOF signal would be dispersive, because in these cases the dynamics of electron distribution, which determines the eventually measured drift mobility, depends on the transit time that is controlled by the applied field or sample thickness. By contrast, if $t_d$ (or $t_0$) is considerably shorted than the TOF duration, the distribution of photogenerated electrons would reach a steady state in a negligible time relative to the transit time, after which the average drift mobility would be invariant. Accordingly, regardless of the transit time, the measured mobility remains invariant; this is non-dispersive transport. $t_d$ (or $t_0$) defines the thermalization time, and $E_D$ (or $E_b$) defines the thermalization depth.

Temperature plays a critical role in determining a TOF transport being non-dispersive or dispersive. Assuming that the transport is initially non-dispersive, $t_d$ (or $t_0$) is far shorter than the total transit time, which remains so regardless of the varying electric field within a given range. As the temperature decreases, $t_d$ (or $t_0$) increases drastically according to Eq. (2). This increase eventually surpasses the increase of total transit time with the decreasing temperature (due to the decrease of the average drift mobility [14]). The results are that $t_d$ (or $t_0$) starts to be comparable with the total transit time, and that the transport starts to be dispersive within the same range of electric field.

It is common that thin films of glow-discharged $a$-Si:H exhibit non-dispersive characteristics in TOF electron measurements around room temperature [14,16-18]. However, different samples transition to dispersive transport at different temperatures. Looking back on the historical studies of $a$-Si:H, two types of $a$-Si:H samples exist which respectively feature high transition temperatures (e.g., ~ 250 K in Ref. [19]) and low transition temperatures (e.g., less than 150 K in Ref. [20]). Under the framework of Case (1), the transition temperature is determined by the steepness of the DOS distribution; a steeper DOS tail leads to a closer distance between $E_C$ and $E_b$, indicative of a lower transition temperature. Historically, the exponential band tail model was adopted as a rough strategy to quantify the tail steepness, in which case the transition temperature is around the characteristic slope of the exponential tail that is typically ~ 220 K to 270 K [18]. This is consistent with the first type of $a$-Si:H samples. By contrast, the second type of $a$-Si:H is better explained under the framework of Case (2), where the drop of capture cross section occurs at a shallow energy in the band tail.

Regardless of being explained by Case (1) or Case (2), as long as a transport is non-dispersive within the temperature range of interest, it was taken for granted by almost all previous literature (see for example, Refs. [9,18,21]) that, according to Eq. (1),

$$\mu_d = \mu_c \frac{N_{ext}}{N_{ext} + N_{loc}}, \quad (5)$$

where $N_{ext}$ and $N_{loc}$ respectively represent, at any moment after the steady state is reached, the number of photogenerated electrons that are in extended states and the number of photogenerated electrons that are in localized states. The ratio between $N_{ext}$ and $N_{loc}$ was believed to be

$$\frac{N_{ext}}{N_{loc}} = \frac{\int_{E_C}^{+\infty} g(E)F(E)dE}{\int_{E_L}^{E_C} g(E)F(E)dE}, \quad (6)$$

where $g(E)$ denotes the DOS distribution, $F(E)$ is a steady Boltzmann distribution determined by Eq. (3), and $E_L$ represents $E_D$ (or $E_b$) depending on which case holds true.

Similar to non-dispersive transport in TOF experiments, another typical steady-state conduction scenario is in DC conduction which is more relevant to the operation of many real electronic devices based on amorphous semiconductors, such as $a$-Si:H thin film transistors (TFTs) used in flat panel displays. For the DC conduction, the conductivity mobility $\mu_a$ (which determines the field effect mobility of the corresponding TFT), in the framework of MTR, takes a similar form to Eq. (5) as [5]

$$\begin{cases} \mu_a = \mu_c \dfrac{n_{ext}}{n_{ext} + n_{loc}} \\ \dfrac{n_{ext}}{n_{loc}} = \dfrac{\int_{E_C}^{+\infty} g(E)f(E)dE}{\int_{E_{L'}}^{E_C} g(E)f(E)dE} \end{cases}, \quad (7)$$

where $n_{ext}$ and $n_{loc}$ respectively represent the density of extended-state electrons and localized-state electrons that are intrinsic to the material, $f(E)$ denotes the Fermi-Dirac distribution, and $E_L'$ represents the energy below which the DOS is trivial.

Revisiting Eqs. (1), (5), (6) and (7), they are all based on the abrupt mobility edge model which embeds two simplifications on the attributes of extended states and localized states (see for example, Refs. [21,22]):

(1) *Simplification of the geometry of states*: any of the extended states is "omnipresent" in space, whereas the existence of localized states in the energy-space diagram is treated as scattered "pointlike" sites.
(2) *Simplification of the uniformity of states*: all states are evenly distributed in space, and the local DOS distribution is spatially invariant.

However, according to the prequel to this paper [23], neither of these two simplifications is accurate. Extended states and localized band tail states possess complex spatial geometries that continuously evolve with energy. Meanwhile, no state is entirely uniform and the local DOS varies spatially. Taking these critical attributes into account, there is a natural doubt on the accuracy of all these widely accepted quantifications.

Considering that MTR transport is still so crucial even for the state-of-the-art development of large-area electronics (LAE), this paper incorporates the insights from the prequel paper and reinvestigates the transport. Section II of this paper starts with an acceptance of Simplification (2) but takes into account the true geometric attributes of extended states and localized band tail states. A uniform and symmetric band edge fluctuation model, idealized from the results in the prequel paper, is established, based on which a rigorous time-domain analysis of the microscopic carrier dynamics is given, leading to a modified MTR quantification. Simplification (2) is then deactivated, the effect of which on the derived results is discussed. To concretize the analysis and derivations in Sec. II, the *a*-Si:H discussed in the prequel paper, which was an intensively studied sample with a set of reliable data, is used in Sec. III as an example case study.

The analyses and findings in this paper are not only theoretically significant, but they also serve as a more reliable guide to practical device engineering. With the unstoppable trend of device downscaling, an accurate understanding of microscopic electronic structures and transport details becomes increasingly important. The sequel to this paper [24] will build on the results in this paper and its prequel [23], and show that the electron mobility of a nanoscale *a*-Si:H device can be ~ 10 times that of a macroscopic *a*-Si:H thin film.

## II. TIME-DOMAIN PROBABLISTIC ANALYSIS OF THE ACTUAL MTR ELECTRON TRANSPORT IN DISORDERED SEMICONDUCTORS EXAMPLIFIED BY *a*-Si:H

In the prequel to this paper [23], the authors have established a framework where the bond angle and bond length distortion in *a*-Si:H generate excess delocalized charges that force the crystalline silicon (*c*-Si) band structure to fluctuate in the energy-space diagram and produce the commonly observed DOS features in and near the CB tail. These excess charges were modeled using a finite-element approach where the basic spatial element, termed a "short-range locality" (SRL), is a cube with a side length ($a_0$) of 3 Å. The corresponding fluctuation of band edge [i.e., conduction band minima (CBM)] induced by the established model is copied here in Fig. 2(a). The band edge of crystalline silicon (*c*-Si) is set as the energy reference (0 eV). Based on this fluctuation, the spatial attribute of extended states and localized band tail states can be specified, several representatives of which are shown in Fig. 2(b). Two facts are clear.

(1) The geometric evolution of states with energy is continuous. Extended states are not necessarily omnipresent, and localized band tail states cannot always be treated as pointlike sites. These two categories of states are distinct only at a macroscopic scale based on whether or not there is a complete path through the states connecting the boundaries of the sample; they do not fundamentally differ from each other from a microscopic viewpoint.
(2) The spatial distributions of both extended states and localized band tail states are not strictly uniform. However, it is worthwhile to note that the lowest spatial frequency in Fig. 2(a) corresponds to a fluctuation length scale of < 100 nm. For typical thin films in micrometer scales, this means that, despite the microscopic nonuniformity, states are still uniform at the macroscale.

In the description of MTR based on the abrupt mobility edge model, the defined mobility edge $E_C$ serves two purposes at the same time. It not only demarcates extended states and localized states, but also distinguishes relaxation (i.e., electron transition above $E_C$) and trapping (i.e., electron transition below $E_C$). Considering Fact (1), although a clear critical energy [$E_{ct}$, shown in Fig. 2(b)] may be defined which demarcates the two types of states, it is difficult to specify relaxation and trapping given the continuous geometric evolution. In the framework of abrupt mobility edge, relaxation is a considerably easier and more rapid process than trapping because trapping to a pointlike state is significantly harder than relaxation between omnipresent extended states. This concept collapses when considering the continuity. Downward electron transition may always be treated as a relaxation process; the concept of trapping, as developed in the abrupt mobility edge model, is essentially still a relaxation process.

Nevertheless, it is noted that the difficulty of downward transition increases as the energy decreases. This is mainly due to the fact that both the DOS and the inoccupation probability [1−$f(E)$] rapidly decreases, the multiplication of which is semi-quantitatively treated as the probability per unit time that an electron transitions downward [see Fig. 2(c)]. Despite the continuous profile of this transition probability, it is acceptable to take an effective discretization approach and define an energy level above which relaxation is deemed "easy" while below which relaxation is deemed "difficult"; the latter now corresponds to "trapping" in the previous sense. It is not straightforward to precisely specify the position of this energy level at present, but it is likely to be higher than $E_{ct}$ simply considering the trend shown in Fig. 2(c); this will be retrospectively discussed later in Sec. III in the context of a specific *a*-Si:H sample from which Fig. 2(c) is obtained. The defined energy level is henceforth termed an *effective mobility edge*, $E_{Ceff}$.

Even with this understanding, the transport problem still seems unsolvable given the complex geometries of extended states and localized band tail states shown in Fig. 2(b). In this regard, it is worthwhile to consider the attribute of macroscopic uniformity revealed in Fact (2). An equivalent analysis could thus be first made based on an ideal band edge fluctuation model which is uniform both macroscopically and microscopically.

In the prequel paper, the authors have revealed that the DOS distribution near the CB tail of *a*-Si:H is determined by the energetic distribution of the local band edges of all SRLs in the material model. This distribution, in the form of a probability density function (PDF), is copied here in Fig. 2(d). Based on the same PDF (and thus ensuring the same DOS distribution), an ideal spatial distribution of local band edges is developed using the method detailed in the Appendix A of this paper. Part of this spatial distribution in a two-dimensional (2D) plane is shown in Fig. 2(e); the length unit $A$ quantifies the distances between the nearest maxima and minima in the fluctuation. The five representative energies specified in Fig. 2(b) are now considered in the ideal band edge fluctuation model, which yield the geometric features of the five corresponding states in the new model; they are shown in Fig. 2(f) using the same colors as in Fig. 2(b). Note that different from the actual model, the ideal model, in essence, is sufficiently large, and so each of the geometric depictions in Fig. 2(f) essentially shows only four repeated units each with a side length of $A$. Merely for the purpose of the analytical discussions in this section, the energy at 0.075 eV is chosen to represent the effective mobility edge ($E_{Ceff}$) of this model. The choice of $A$ is random on condition that $A$ is short relative to the typical range of electron thermal motion at $E_{Ceff}$; this will be discussed again shortly.

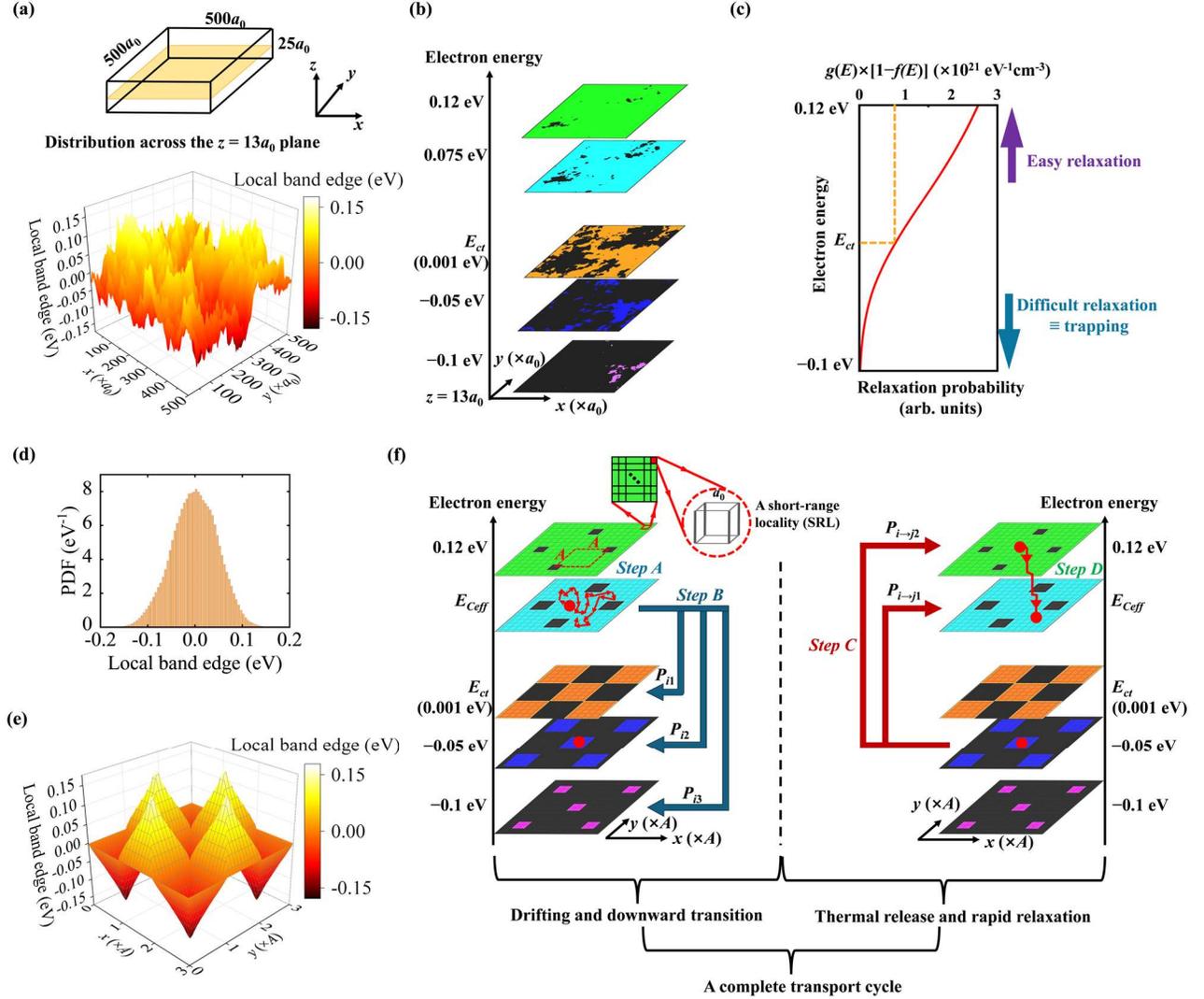

FIG. 2. Analysis of electron transport in amorphous semiconductors based on the attributes of extended states and localized band tail states revealed in the prequel paper [23]. (a) The fluctuation of local band edge in the middle layer of the hydrogenated amorphous silicon ($a$-Si:H) model developed in the prequel paper. $a_0$ denotes the side length of short-range localities (SRLs) which serve as the basic modeling elements. The band edge of crystalline silicon ($c$-Si) is set as the energy reference, 0 eV. (b) Representative two-dimensional (2D) geometries of extended states and localized states at five distinct energies. Different colors indicate different states. Electrons are prohibited from the black regions. $E_{ct}$ denotes the critical energy that demarcates extended states and localized states. (c) Semiquantitative reflection of the ease of downward relaxation. This is calculated based on the multiplication of Fermi-Dirac distribution [$f(E)$] and the density of states (DOS) distribution [$g(E)$] in the prequel paper [23]. (d) The energetic distribution of the band edge fluctuation in (a) quantified using the probability density function (PDF) (reproduced from the prequel paper [23]). (e) An ideal and microscopically uniform band edge fluctuation model that matches the PDF in (d). Shown here is the band edge fluctuation across a 2D plane that spans four complete unit cells, each being a cube with a side length of $A$. (f) Transport analysis targeting the ideal band edge fluctuation model in (e). Under the discretization treatment, relaxation above the effective mobility edge ($E_{Ceff}$) is easy whereas downward transitions below $E_{Ceff}$ is relatively difficult. The inset illustrates a SRL within a unit cell. The entire transport is decomposed into numerous transport cycles that are mutually independent. Each cycle contains four steps.

Similar to the description in Fig. 1, the evolution of a representative electron carrier in a DC conduction is considered in this new system, which decomposes into several distinct steps that repeat till the electron completely transit through the sample. Non-dispersive TOF conduction is highly similar and will be discussed later in Sec. II F.

### A. Interaction with Deeper Electronic States Whilst Remaining at $E_{Ceff}$

Under the discretization treatment, before further transitioning downward, electron carriers predominantly remain at $E_{Ceff}$ rather than at higher energies due to the rapid relaxation. Assuming that the average thermal velocity of electrons is $v_{th}$, an electron at $E_{Ceff}$ will travel a distance of $v_{th}\tau_0$ per unit time $\tau_0$ in a zig-zagged manner whilst remaining at $E_{Ceff}$; this is illustrated by *Step A* in Fig. 2(f). Note that the zig-zagged electron path in disordered semiconductors is not only due to scattering from atomic vibration but is also due to the disorder-induced scattering associated with the band fluctuation, plus the scattering from the black forbidden regions in the figure.

As annotated in the prequel paper [23] and shown in Fig. 3, due to the fact that momentum ceases to be a good quantum number in disordered solids, an extended state or a localized state is defined by energy and should be understood as an ensemble of *basic electronic states* at the same energy; these basic electronic states are contributed by different SRLs and may correspond to different momenta.

In time $\tau_0$, the electron can effectively interact with the basic electronic states at a deep energy $E_i$ if these basic electronic states exist in a quasicylindrical region that surrounds the electron path; the volume of this quasicylindrical region is $v_{th}\tau_0\bar{\sigma}(E_{Ceff}, E_i)$. Here, $\bar{\sigma}(E_{Ceff}, E_i)$ is an average capture cross section that reflects the spatial range within which an electron can be captured from $E_{Ceff}$ into the energy level $E_i$ at a SRL. The term "average" is used because little is known on the attributes of the capture cross sections, which might be associated with the local momentum that is SRL-specific. $\bar{\sigma}(E_{Ceff}, E_i)$ therefore represents an average value which, strictly speaking, depends on the specific path that the electron takes.

Among these nearby basic electronic states that surround the electron path during $\tau_0$, the number of unoccupied basic electronic states in the energy interval $\Delta E$ around $E_i$ is

$$[1 - f(E_i)]g_L(E_i)\Delta E v_{th}\tau_0\bar{\sigma}(E_{Ceff}, E_i), \quad (8)$$

where $g_L$ represents the local DOS in the vicinity of the electron path. Importantly, as long as the length unit $A$ is short relative to the spatial range of electron thermal motion at $E_{Ceff}$, $g_L$ would be equal to the macroscopic DOS ($g$). Similarly, in this case, the average capture cross section $\bar{\sigma}(E_{Ceff}, E_i)$ would be independent of the specific path taken by the electron and would be equal to its counterpart defined at the macroscale which is now denoted as $\sigma(E_{Ceff}, E_i)$. Without losing generality, it is admitted that $\sigma(E_{Ceff}, E_i)$ can exhibit a complex energetic dependence.

The longer the time that the electron remains at $E_{Ceff}$, the more unoccupied basic electronic states it interacts with, and the higher the probability that this electron will start to transition downward. Therefore, a duration termed free time ($t_f$) is defined which specifies the duration that the electron remains at $E_{Ceff}$ before downward transition occurs. From a statistic viewpoint, the most probable value of $t_f$, $\bar{t}_f$, is quantified by taking the total number of unoccupied basic electronic states with which the electron has interacted as unity. In another word,

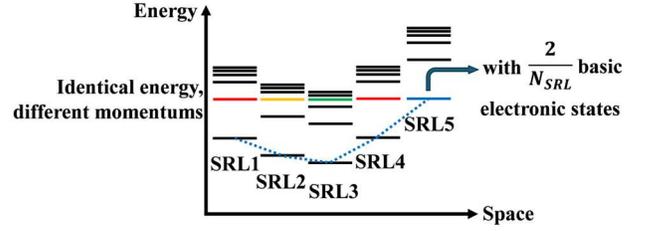

FIG. 3. Illustration of basic electronic states in the energy-space diagram. Five short-range localities (SRLs) are displayed in a one-dimensional (1D) setting, each contributing a fraction ($2/N_{SRL}$) of one basic electronic state to an extended state at a certain energy. The spin degeneracy is considered. $N_{SRL}$ denotes the total number of SRLs divided in the material. The $2/N_{SRL}$ basic electronic states at different SRLs possess different local momentums. A complete basic electronic state holds a maximum of one electron. The dashed blue curve depicts the fluctuation of local conduction band edge.

$$\sum_{p=1}^{m}\{\bar{t}_f v_{th}\sigma(E_{Ceff}, E_p)[1 - f(E_p)]g(E_p)\Delta E\} = 1, \quad (9)$$

assuming that there are totally $m$ discretized energy levels below $E_{Ceff}$, and that $A$ is short as discussed earlier. Here, $p$ indexes energy levels below $E_{Ceff}$ to avoid being confused with $i$.

### B. Downward Transition ("Trapping")

On average, downward transition (i.e., trapping, in the previous sense) occurs after the electron remains at $E_{Ceff}$ for a duration of $\bar{t}_f$. However, transitions to different energy levels are not equally probable. The more the electron interacts with unoccupied basic electronic states at a certain energy, the higher chance that the electron will transition to this energy level. According to Eq. (8), the probability of transitioning to $E_i$ is simply a conditional probability quantified by

$$P_i = \frac{[1 - f(E_i)]g(E_i)\Delta E v_{th}\sigma(E_{Ceff}, E_i)}{\sum_{p=1}^{m}\{[1 - f(E_p)]g(E_p)\Delta E v_{th}\sigma(E_{Ceff}, E_p)\}}. \quad (10)$$

Assume that the electron exactly transitions to $E_i$, which is illustrated in Fig. 2(f) by the red sphere at −0.05 eV as an example. The electron then remains at $E_i$ for a certain duration which is known as the thermal release time, $t_r$, before being released.

Now, given the continuity, it should be realized that the above description, following the picture of abrupt mobility edge model, does not entirely reflect the actual situation. In reality, there could be a more complex picture of electron transitions: the electron does not have to directly transition to $E_i$; instead, it can first transition to a shallower energy from $E_{Ceff}$ before further transitioning downward to $E_i$. This is defined as an indirect transition. The same applies to the duration while the electron remains at $E_i$. There is a chance that the electron will further transition downward before being thermally released.

Such a complex picture was not considered in the abrupt mobility edge model. In that model, transitions between localized states were highly difficult compared with the direct transition from the omnipresent mobility edge, because localized states were treated as pointlike sites whose wavefunctions overlap with each other very trivially. If, in any case, the transitions between localized states occur, that should be categorized to hopping transport which is excluded from the MTR regime and which is only observable at a very low temperature [4].

Although the indirect transitions are not prohibited in the current continuity model, it worths investigating the likelihood of the occurrence of indirect transitions. In this regard, the thermal release process, which will be shortly mentioned, plays a critical role. Similar to the understanding in Refs. [11,15], in this paper, the average time for the electron at $E_i$ to be thermally released to anywhere above $E_{Ceff}$ may be roughly quantified by

$$t_r(E_i) \sim \nu^{-1} \exp\left(\frac{E_{Ceff} - E_i}{kT}\right). \quad (11)$$

Clearly, if the electron transitions from $E_{Ceff}$ to a shallow state, it is easily and rapidly released. Given that downward transition is generally difficult below $E_{Ceff}$, the predominant case should be that the electron does not have a chance to further transition downward before it is released. Although the average thermal release time increases exponentially with a decreasing energy, the probability of downward transition typically shrinks more rapidly in the meantime in accordance with the decrease of DOS, the increase of occupation [and possibly, the decrease of capture cross section (e.g., Ref. [19])]. The result should be that thermal release outperforms continuous downward transition in most scenarios, such that neglecting the complex indirect transitions remains a good approximation even without the aid of the abrupt mobility edge model. Consequently, Eq. (10) remains accurate and acceptable.

### C. Thermal Release

As the electron at $E_i$ gains sufficient energy from the solid, thermal release occurs, which is illustrated by *Step C* in Fig. 2(f). Similar to the downward transition from $E_{Ceff}$, the probability per unit time that the electron at $E_i$ is released to a shallow energy $E_k$ is correlated with the density of unoccupied basic electronic states at $E_k$ within the spatial range where the electron remained at $E_i$. Moreover, an emission coefficient $e(E_i, E_k)$ which indicates a preference for low energies should also be introduced similar to Ref. [25]. Accordingly, this unit-time probability is expressed by

$$P_{ik} \propto [1 - f(E_k)] g_L(E_k) \Delta E e(E_i, E_k). \quad (12)$$

Different from Eqs. (8) and (10) where the macroscopic DOS is used, the DOS used in Eq. (12) is the local DOS specific to the spatial range where the electron remained before thermal release. Based on the band edge fluctuation model in the prequel paper [23], the deeper the electron was before thermal release, the higher this local DOS would be; by contrast, the shallower the electron was before thermal release, the closer to the macroscopic DOS this local DOS would be.

To incorporate this dependence on $E_i$, a DOS scaling factor $S(E_i)$ is introduced which transforms Eq. (12) to

$$P_{ik} \propto [1 - f(E_k)] g(E_k) S(E_i) \Delta E e(E_i, E_k), \quad (13)$$

by which the same macroscopic DOS ($g$) can be used in any case. Combining $S(E_i)$ with $e(E_i, E_j)$, an effective emission coefficient $e'(E_i, E_j) = S(E_i) \times e(E_i, E_j)$ is defined, so that

$$P_{ik} \propto [1 - f(E_k)] g(E_k) \Delta E e'(E_i, E_k), \quad (14)$$

Electrons below $E_{Ceff}$ always have chances to be released back to states above $E_{Ceff}$, and then rapidly relax to $E_{Ceff}$ according to the discretization treatment discussed earlier. Focusing on the representative electron in this section, it will be eventually released to the $j^{\text{th}}$ level ($E_j$) above $E_{Ceff}$. However, similar to the case of downward transition, the thermal release may either take a direct form or an indirect form. In a direct thermal release event, the electron at $E_i$ gains sufficient energy from the solid and is released to $E_j$ in one step. By contrast, in a typical indirect release event the electron is first released to a shallow state ($E_u$) below $E_{Ceff}$, followed by a secondary release from $E_u$ to $E_j$.

Such an indirect release event is again not considered in the conventional abrupt mobility edge model, because only extended states above the mobility edge are omnipresent whereas other localized states are pointlike sites that are unlikely to present at the exact spatial locality where the electron was trapped. However, in reality, due to the continuity, the state at $E_i$ is spatially enclosed in any higher state, allowing the onset of release to any higher state that is not necessarily above $E_{Ceff}$.

Unlike indirect downward transition which is highly rare due to the drastically decreasing DOS as well as the counteraction of thermal release, indirect release is expected to be slightly more common. Despite the fact that a higher energy supports more unoccupied basic electronic states, it meanwhile consumes more thermal energy for the electron to be released there than to a lower energy. This paper therefore specially takes indirect release into consideration, and for simplicity, focuses are given to the case where only one intermediate state is involved because those including multiple intermediate states are less probable given the aforementioned balance between the number of unoccupied basic electronic states and thermal energy consumption.

In this framework, the probability of being released from $E_i$ to $E_j$, $P_{i \to j}$, is a result of the probability addition rule:

$$P_{i \to j} = P_{ijD} + P_{ijI}, \quad (15)$$

where $P_{ijD}$ and $P_{ijI}$ respectively denote the probability of direct release and the probability of indirect release from $E_i$ to $E_j$. According to Eq. (14), the former is a conditional probability which, taking into account all possible single-step release possibilities, quantifies the relative likelihood that the electron is being released specifically to $E_j$. Therefore,

$$P_{ijD} = \frac{[1 - f(E_j)] g(E_j) \Delta E e'(E_i, E_j)}{\sum_{p=1}^{i-1}\{[1 - f(E_p)] g(E_p) \Delta E e'(E_i, E_p)\} + \sum_{l=1}^{n}\{[1 - f(E_l)] g(E_l) \Delta E e'(E_i, E_l)\}}. \quad (16)$$

Here, the summations in the denominator starts from $E_i$ and covers the entire CB that is discretized into $n$ quasicontinuous energy levels. Energies below $E_{Ceff}$ are indexed by $p$ [downward direction, following the notation in Eq. (10)], while energies above $E_{Ceff}$ are indexed by $l$ (upward direction); they are independent indices not to be confused with $j$. For simplicity, the entire denominator of Eq. (16) is henceforth denoted as $\xi$. The indirect release probability $P_{ijI}$ first follows a further addition rule that considers various possible intermediate energies; each of these sub-categories then follows the multiplication rule reflecting the probability of the second release to $E_j$. Therefore,

$$P_{ijI} = \sum_{u=1}^{i-1}\left[\frac{[1-f(E_u)]g(E_u)\Delta E\,e'(E_i,E_u)}{\xi}\frac{[1-f(E_j)]g(E_j)\Delta E\,e'(E_u,E_j)}{\sum_{l=1}^{n}\{[1-f(E_l)]g(E_l)\Delta E\,e'(E_u,E_l)\}}\right] \quad (17)$$

Here, the fact that indirect release with multiple intermediate energies are not considered has been treated as a preset condition, so the second summation in the denominator starts from $E_{Ceff}$ (i.e., $l = 1$) to express the conditional probability that the second release brings the electron from $E_u$ to $E_j$ specifically. For the convenience of what follows, the $(i-1)$ terms in Eq. (17) are respectively denoted as $P_{ijIu}$, with $u$ between 1 and $(i-1)$.

The thermal release time $t_r$ defined in Sec. II B quantifies the duration that the electron remains at $E_i$. Without losing generality, this thermal release time should be treated as being dependent on both $E_i$ and the energy level the electron is firstly released to. Therefore, in the case that the electron is directly released from $E_i$ to $E_j$, the duration that the electron remains at $E_i$ is denoted as $t_r(E_i, E_j)$. Similarly, in the case that the electron is released from $E_i$ to $E_u$ and then to $E_j$, the electron would remain at $E_i$ for $t_r(E_i, E_u)$ and then remain at the intermediate level $E_u$ for $t_r(E_u, E_j)$ before reaching $E_j$.

Thermal equilibrium is established in the steady-state DC conduction. Excluding statistical fluctuation, starting from any moment, there are $N_e = f(E_i)g(E_i)\Delta E$ electrons per unit volume occupying the basic electronic states within the energy interval $\Delta E$ around $E_i$ (this energy interval is henceforth deemed as a single level for the ease of expression). Although electrons are indistinguishable, it provides analytical convenience and clarity by labeling these electrons using an index $\alpha$ with $\alpha \in [1, N_e]$. Accordingly, the $\alpha^{th}$ of these $N_e$ electrons is referred to as "electron $\alpha$". As illustrated in Fig. 4(a) in the time domain, "electron $\alpha$" will be thermally released to a specific higher energy at some point afterwards, but this release event will be immediately followed by the transition of a new electron (labeled as "electron $\alpha,1$") to $E_i$ in the unit volume so as to maintain thermal equilibrium. This new electron may come to $E_i$ either through a transition from $E_{Ceff}$ or through a transition from an energy level lower than $E_i$ as a result of an indirect thermal release. These two scenarios are respectively illustrated by the blue and green arrows in Fig. 4(b). The same release and immediate compensation process applies to "electron $\alpha,1$" as well as any of its successors which are each indexed as "electron $\alpha, \beta$" with $\beta > 1$. Each of these electrons remains at $E_i$ for a specific duration which is quantified by the energy-dependent release time. As shown in Fig. 4(a), from a temporal perspective, there is neither overlap nor gap between these durations for the same $\alpha$, so the sum of the durations of all the electrons that have ever existed at $E_i$ throughout the measurement duration $T_m$ is $T_m N_e$ for the unit volume.

From another perspective, all these electrons can be classified according to which energy levels above $E_{Ceff}$ they are specifically released to in the end (regardless of being in a direct or indirect manner). These energy levels above $E_{Ceff}$ are indexed by $j$. Suppose that the emission rate at $E_i$ is $R(E_i)$, which quantifies the total number of electrons in the unit volume that are released away from $E_i$ per unit time. Apparently, this emission rate includes both direct and indirect release. Among these released electrons, a fraction ($P_{ijD}$) are directly released to $E_j$ according to Eq. (16), whereas a fraction ($P_{ijI}$) are indirectly released to $E_j$ via a certain intermediate energy according to Eq. (17). For the former case, this part of electrons share the same thermal release time of $t_r(E_i, E_j)$. By contrast, for the latter case, the duration that these electrons remain at $E_i$ depends on the intermediate energy. Among the electrons in the latter case, a fraction $(P_{ijIu}/P_{ijI})$ remain at $E_i$ for a duration of $t_r(E_i, E_u)$. Hence, the sum of the durations of all electrons that have ever existed at $E_i$ during $T_m$ may be alternatively quantified from the second perspective, which establishes an equation that

$$T_m f(E_i) g(E_i) \Delta E = T_m R(E_i) \sum_{j=1}^{n} \left\{ t_r(E_i, E_j) P_{ijD} + \sum_{u=1}^{i-1} [t_r(E_i, E_u) P_{ijIu}] \right\}. \quad (18)$$

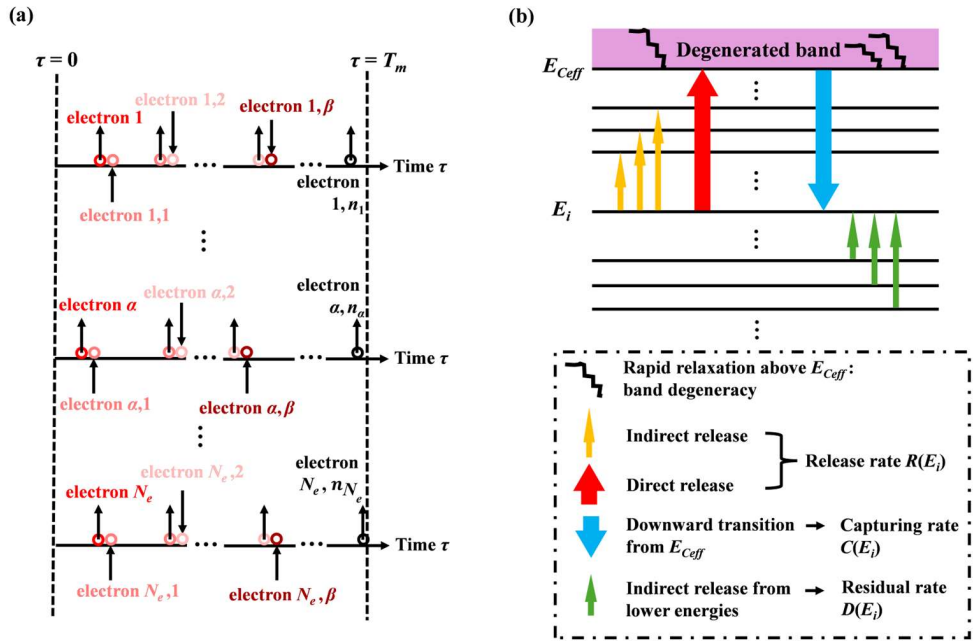

FIG. 4. Analysis of the microscopic dynamics under thermal equilibrium. (a) Time-domain analysis of electron flows at a specific energy level $E_i$. The time scale under investigation is the DC measurement duration $T_m$. Each of the initial $N_e$ electron carriers intrinsic to a unit volume leads a "temporal chain" on which the release of a former electron is immediately compensated by another incoming electron which may either be captured from the effective mobility edge ($E_{Ceff}$) or be released from a deeper energy. For ease of analysis, all the electrons that have ever existed at $E_i$ during $T_m$ are labeled according to the rule in text. (b) Illustration of the balance under thermal equilibrium. Because relaxation above $E_{Ceff}$ is rapid under the discretization treatment adopted in this paper, states above $E_{Ceff}$ can be treated as a whole and as a degenerated band.

Focus now moves to the evaluation of $R(E_i)$, the emission rate at $E_i$. Under thermal equilibrium, the principle of detailed balance is maintained. Focusing on the energy level $E_i$, as shown in Fig. 4(b), the balance is among four categories of electron transitions: (i) direct release of electrons from $E_i$; (ii) indirect release of electrons from $E_i$; (iii) indirect release of electrons from deeper energies to $E_i$; (iv) capturing of electrons from $E_{Ceff}$. For the fourth category, it should be noted that, although electrons may be released to energies that are higher than $E_{Ceff}$, these released electrons are eventually captured from $E_{Ceff}$ rather than the higher energies. This is because, under the discretization treatment in this paper, electrons at any higher energies than $E_{Ceff}$ would rapidly relax to $E_{Ceff}$ before they can be captured by any state below $E_{Ceff}$. Therefore, all energy levels above $E_{Ceff}$, as a whole, may be treated as a degenerated band, and capturing of electrons from this band always occurs at the bottom of the band (i.e., at $E_{Ceff}$). See the note in [26] as well. The capture rate from this degenerated band to $E_i$ per unit volume is

$$C(E_i) = \left[ \sum_{j=1}^{n} f(E_j) g(E_j) \Delta E \right] v_{th} \sigma(E_{Ceff}, E_i) [1 - f(E_i)] g(E_i) \Delta E, \quad (19)$$

according to Eq. (8). Different from the conventional abrupt mobility edge model, because of the existence of indirect release, $R(E_i)$ is not strictly equal to $C(E_i)$. Instead,

$$R(E_i) = C(E_i) + D(E_i), \quad (20)$$

where, as illustrated in Fig. 4(b), $D(E_i)$ denotes the residual rate supplied by the indirect release from deeper energies. Meanwhile, it should be noted that the duration in the secondary summation of Eq. (18) is related to $t_r(E_i, E_u)$ only but not the complete thermal release time that the electrons will take in order to be eventually released to $E_j$ above $E_{Ceff}$. For the latter, the total thermal release time is $t_r(E_i, E_u) + t_r(E_u, E_j)$. Now, reasonably assume that indirect release, although more common compared with indirect downward transition, is still less frequent than direct release. In this case, the probability $P_{ijIu}$, as well as $D(E_i)$ may be treated as perturbations. As a first-order approximation, Eq. (18) can be approximated by

$$T_m f(E_i) g(E_i) \Delta E = T_m [R(E_i) - D(E_i)] \sum_{j=1}^{n} \left[ t_r(E_i, E_j) P_{ijD} + \sum_{u=1}^{i-1} \{ [t_r(E_i, E_u) + t_r(E_u, E_j)] P_{ijIu} \} \right]. \quad (21)$$

which means,

$$f(E_i) g(E_i) \Delta E = C(E_i) \sum_{j=1}^{n} \left[ t_r(E_i, E_j) P_{ijD} + \sum_{u=1}^{i-1} \{ [t_r(E_i, E_u) + t_r(E_u, E_j)] P_{ijIu} \} \right]. \quad (22)$$

### D. Relaxation and Drifting

As illustrated in Fig. 2(f), after the electron eventually reaches an energy higher than $E_{Ceff}$ through either direct or indirect thermal release, the electron will rapidly relax to $E_{Ceff}$ according to the discretization treatment. The electron will then restart drifting as is in *Step A*, through which it interacts with the basic electronic states at various energies and at various SRLs. The relaxation process is deemed sufficiently rapid such that it is trivial in the time domain compared with the subsequent drifting at $E_{Ceff}$. This is the fundamental reason why the capture cross sections in Eqs. (8), (9) and (10) are always correlated with $E_{Ceff}$ rather than any higher energies, and the reason why energy levels above $E_{Ceff}$ can be treated as a single degenerated band as mentioned earlier.

### E. Time-domain Drift Mobility Quantification

Assuming that the material is sufficiently large (e.g., considering a typical thin film sample in the micrometer scales), *Step A* to *Step D* will continue in sequence for numerous times until the representative electron reaches the electrode at the boundary of the material. The average electron mobility of this material is therefore quantified by the average drift mobility of this representative electron because the electronic structure in the material is uniform and different electron carriers would end up exhibiting statistically identical average mobility.

To quantify this average drift mobility, the entire migration process of the electron is decomposed into individual cycles. Any single cycle is made up of *Step A* to *Step D* which proceed in sequence. "Cycle $i, j$" is defined as a process that begins from the moment the electron is captured from $E_{Ceff}$ into the $i^{th}$ energy level below $E_{Ceff}$ and ends at the moment the same electron is about to be re-captured again from $E_{Ceff}$. Within this cycle, the electron is released (either directly or indirectly) to the $j^{th}$ level above $E_{Ceff}$ before its rapid relaxation down to $E_{Ceff}$. Exactly because of this rapid relaxation, every cycle starts with the capturing of the electron from $E_{Ceff}$, and so each cycle is deemed to be fully independent of its predecessors, so the entire electron migration process is now merely a *random combination* of all such mutually independent cycles.

The definition in Eq. (1) states that the drift mobility is controlled by the cumulative free time and the cumulative trapping time. Now, in the framework of this paper, the cumulative free time is simply $\bar{t}_f$ multiplied by the total number of cycles which is henceforth denoted as $N_{total}$. The cumulative trapping time, in this paper, refers to the cumulative time that the electron spends below $E_{Ceff}$, and is therefore a sum of all thermal release time from every cycle. In this regard, it is worthwhile to note that the word "trapping" is not used. This is because when the electron is below $E_{Ceff}$, it is not necessarily highly confined in space as is depicted by the usual definition of trapping. Different from trapping at a localized defect state that is associated with an infinitesimal physical entity such as a dangling bond (DB), remaining at a localized band tail state does not prohibit electron motion within a certain spatial range; as shown in Fig. 2(b) and (f), localized band tail states, associated with structural disorder, do not exhibit extreme confinement unless being at very deep energies. Moreover, if as assumed, $E_{Ceff} > E_{ct}$, the electron may sit in an extended state before being released to above $E_{Ceff}$, in which case it is even less suitable for this phase to be described as trapping. An important implication of the above understanding is that the electron is not frozen whilst it is not at or above $E_{Ceff}$; this is in stark contrast to the abrupt mobility edge model which treats electrons as being completely immobile below the mobility edge. It is therefore necessary to quantify the average drift mobility in a more reliable way than Eq. (1).

Considering the geometric features of states exhibited in either Fig. 2(b) or Fig. 2(f), deep states typically possess satellite sites that are highly confined in space; meanwhile, as semiquantitatively quantified by Eq. (11), an electron in a deep state requires a longer time to be thermally released compared with those in shallow states. Therefore, a deep-state electron is expected to experience frequent scattering by the state boundaries as illustrated in Fig. 5(a), whilst it remains at the deep energy. This scattering results in a high degree of localization such that the electron motion does not contribute to net drift current.

Treating deep-state electrons as being frozen is therefore still valid. By contrast, the spatial spread of a shallow state can be significant; meanwhile, the thermal release time of a shallow-state electron can be short. Therefore, as shown in Fig. 5(b) and (c), it is likely that under the applied field, the spatial location where the electron is just released from the state differs noticeably from the spatial location where the electron was just captured; this leads to effective conduction even at a localized state that was previously believed to only confine electrons.

To systematically investigate the effect of conduction below $E_{Ceff}$, a concept termed *local mobility* ($\mu_{Cl}$) is defined, which specifies the transient drift mobility exhibited by an electron at a state below $E_{Ceff}$, before the onset of any boundary scattering. As illustrated in Fig. 5(d), the local mobility is expected to be energetically invariant if the fluctuation synchronicity detailed in the prequel [23] holds true within the energy range of interest here. In contrast, the conventional "energy-specific mobility" is a global average quantity. Although localized-state electrons can move locally, they cannot carry net current throughout the material, so within the regime of MTR transport (where hopping is not considered), the global mobility associated with any localized state (i.e., below $E_{ct}$) is strictly zero.

Considering the different extent of localization at different energies as well as their thermal release time, a factor termed *drift efficiency*, $\alpha(E)$, is defined, which specifies the proportion of time that the motion of an electron carries net current whilst this electron remains at an energy $E$. With these new insights, and considering the statistical random combination of different cycles, the average drift mobility, like the form of Eq. (1), is

$$\mu_a = \frac{\mu_0 \bar{t}_f N_{total} + \mu_{Cl} \sum_{i=1}^{m} \sum_{j=1}^{n} \left[ \alpha(E_i) t_r(E_i, E_j) N_{ijD} + \sum_{u=1}^{i-1} \{ [\alpha(E_i) t_r(E_i, E_u) + \alpha(E_u) t_r(E_u, E_j)] N_{ijIu} \} \right]}{\bar{t}_f N_{total} + \sum_{i=1}^{m} \sum_{j=1}^{n} \left[ t_r(E_i, E_j) N_{ijD} + \sum_{u=1}^{i-1} \{ [t_r(E_i, E_u) + t_r(E_u, E_j)] N_{ijIu} \} \right]}, \quad (23)$$

where $\mu_0$ denotes the global mobility of free electrons at the effective mobility edge $E_{Ceff}$, $N_{ijD}$ and $N_{ijIu}$ respectively represent the number of cycles in which the electron is directly released from $E_i$ to $E_j$ and the number of cycles in which the electron is indirectly released from $E_i$ to $E_j$ via a specific intermediate energy level at $E_u$. Given that all cycles are mutually independent, $N_{ijD}$ and $N_{ijIu}$ can be quantified by their respective probabilities using Eqs. (10), (16) and (17), following the multiplication rule,

$$\begin{cases} N_{ijD} = N_{total} P_i P_{ijD} \\ N_{ijIu} = N_{tota} \, P_i P_{ijIu} \end{cases}. \quad (24)$$

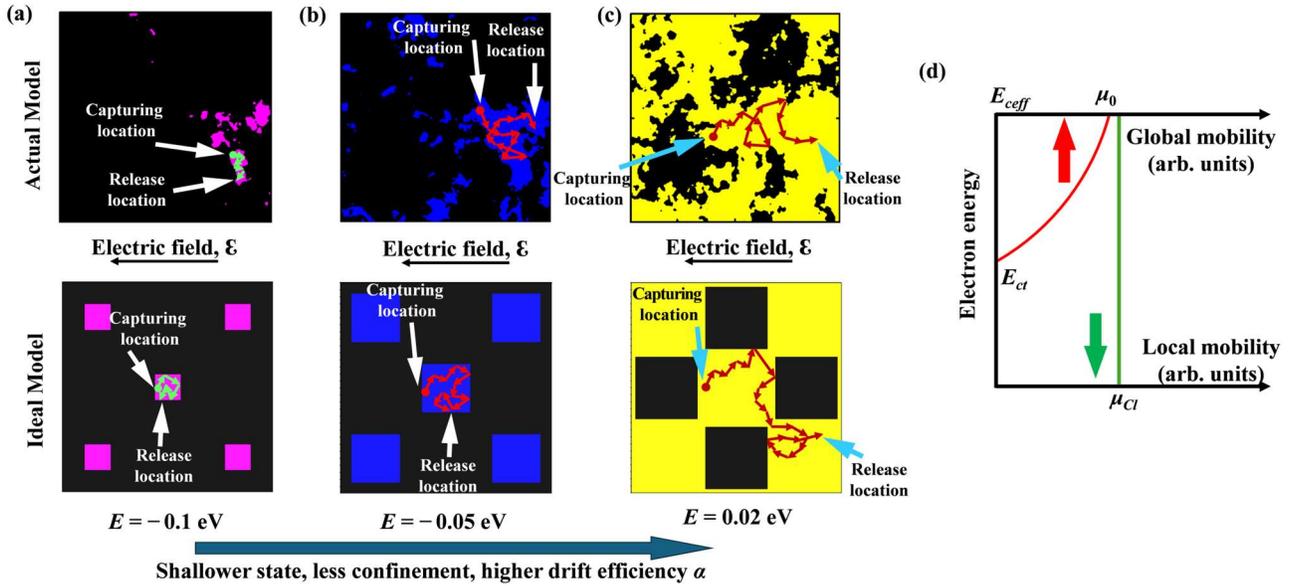

FIG. 5. Microscopic electron motion and the resultant mobility. (a) – (c) Illustrations of the motion of a representative electron whilst it remains at different energies. From (a) to (c), the state energies progressively increase in accordance with the energy values in Fig. 2. The cases in both the actual model (top panels) and the ideal model (bottom panels) are shown. The direction of electric field is labeled in the middle. Once the electron transitions from $E_{Ceff}$ to one of these example states, it continues thermal motion, but the extent of confinement at these different states means that the localization of electron is weaker at shallower energies. Before being thermally released, the electron would exhibit more noticeable net drift motion at shallower energies than at deeper energies, corresponding to a higher drift efficiency $\alpha$. (d) Schematics of the energy dependence of global mobility (the usual definition of mobility, macroscopic quantity) and local mobility (microscopic quantity). As discussed in the prequel paper [23], the global mobility continuously drops to zero as the energy decreases to the critical energy $E_{ct}$. By contrast, the local mobility is approximately constant ($\approx \mu_{Cl}$) within a certain energy range. The global mobility at $E_{Ceff}$ is denoted as $\mu_0$.

This is on condition that $N_{total}$ is sufficiently large to allow the exhibition of statistical nature. In this case, Eq. (23) would be uncorrelated with $N_{total}$ and is determined by probabilities. Considering that electrons still experience boundary scattering even at $E_{Ceff}$, as illustrated in Fig. 2(f), $\mu_0 < \mu_{Cl}$ is expected [as indicated in Fig. 5(d)], but the difference should not be significant if $E_{ceff}$ is sufficiently high.

Equation (22), being a result of thermal equilibrium, can significantly reduce the complexity of Eq. (23). Combined with Eqs. (19), (10) and (9), the complex summation term in the denominator of Eq. (23) is replaced by

$$\sum_{i=1}^{m} P_i \frac{f(E_i) g(E_i) \Delta E}{C(E_i)}$$

$$= \sum_{i=1}^{m} \left( \left[ \frac{[1-f(E_i)] g(E_i) \Delta E v_{th} \sigma(E_{Ceff}, E_i)}{\sum_{p=1}^{m} \{[1-f(E_p)] g(E_p) \Delta E v_{th} \sigma(E_{Ceff}, E_p)\}} \right] \left[ \frac{f(E_i) g(E_i) \Delta E}{v_{th} \sigma(E_{Ceff}, E_i) [1-f(E_i)] g(E_i) \Delta E \sum_{j=1}^{n} [f(E_j) g(E_j) \Delta E]} \right] \right)$$

$$= \sum_{i=1}^{m} \frac{f(E_i) g(E_i) \Delta E}{\sum_{p=1}^{m} \{[1-f(E_p)] g(E_p) \Delta E v_{th} \sigma(E_{Ceff}, E_p)\} \sum_{j=1}^{n} [f(E_j) g(E_j) \Delta E]}$$

$$= \bar{t}_f \frac{\sum_{i=1}^{m} f(E_i) g(E_i) \Delta E}{\sum_{j=1}^{n} f(E_j) g(E_j) \Delta E}. \quad (25)$$

The attribute of the drift efficiency $\alpha(E)$ needs to be understood before the second complex summation in the numerator of Eq. (23) can be simplified. It is difficult now to accurately quantify $\alpha(E)$, but semiquantitative insights are easily obtained.
(1) $\alpha(E)$ is monotonically increasing with energy, meaning that an electron at a higher energy tends to spend a larger proportion of time undergoing effective drift and contributing to the net drift conduction.
(2) $\alpha(E)$ is between zero to unity. It saturates at (or near) unity for high energies.
(3) $\alpha$ depends on the applied field ($\mathcal{E}$) [i.e., $\alpha = \alpha(E, \mathcal{E})$]. Assuming that the thermal release time is unaffected by the change of electric field, under a higher electric field, the boundary scattering would occur earlier, thus reducing $\alpha$.
(4) $\alpha$ depends on temperature [i.e., $\alpha = \alpha(E, \mathcal{E}, T)$]. Under the same electric field, the local drift velocity of a localized-state electron is unchanged. As the temperature rises, the thermal release time of this electron decreases [in an exponential manner if roughly following Eq. (11)]), thus increasing the proportion of time in which the electron drifts effectively and increasing $\alpha$.

Based on Attribute (1) and (2), and given the fact that, as assumed earlier, indirect thermal release serves as a perturbation, $\alpha(E_u)$ in the second summation term of the numerator of Eq. (23) may be replaced by $\alpha(E_i)$, which only leads to a very trivial underestimation of the actual average drift mobility. Under such a first-order approximation, the summation becomes

$$\sum_{i=1}^{m} \alpha(E_i) \sum_{j=1}^{n} \left[ t_r(E_i, E_j) N_{ij} + \sum_{u=1}^{i-1} \{[t_r(E_i, E_u) + t_r(E_u, E_j)] N_{ijIu}\} \right] = \bar{t}_f \frac{\sum_{i=1}^{m} \alpha(E_i) f(E_i) g(E_i) \Delta E}{\sum_{j=1}^{n} f(E_j) g(E_j) \Delta E}, \quad (26)$$

according to the derivation in Eq. (25). Now, under the framework of discretization treatment in this paper, states above $E_{ceff}$ are defined as *effective extended states*, whereas states below $E_{ceff}$ are defined as *effective localized states*; they are separated by the effective mobility edge $E_{ceff}$. Therefore, based on Eqs. (25) and (26), Eq. (23) is concisely expressed as

$$\mu_a = \frac{\mu_0 n_{ext,eff} + \mu_{Cl} n_{loc,eff}'}{n_{ext,eff} + n_{loc,eff}}, \quad (27)$$

where $n_{ext,eff}$ and $n_{loc,eff}$ respectively express the density of electrons at effective extended states (termed *effective extended-state electrons*) and those at effective localized states (termed *effective localized-state electrons*). They are demarcated by $E_{Ceff}$.

$$\begin{cases} n_{ext,eff} = \sum_{j=1}^{n} [f(E_j) g(E_j) \Delta E] \\ n_{loc,eff} = \sum_{i=1}^{m} [f(E_i) g(E_i) \Delta E] \end{cases}. \quad (28)$$

$n_{loc,eff}'$ is an inner product of the form

$$n_{loc,eff}' = \sum_{i=1}^{m}[\alpha(E_i)f(E_i)g(E_i)\Delta E]. \quad (29)$$

Given Eqs. (27) and (29), it is feasible to define an effective extended-state mobility such that

$$\mu_{Ceff} = \mu_0 + \mu_{Cl}\frac{n_{loc,eff}'}{n_{ext,eff}}. \quad (30)$$

The average drift mobility can therefore be re-expressed as

$$\mu_a = \mu_{Ceff}\frac{n_{ext,eff}}{n_{ext,eff} + n_{loc,eff}}. \quad (31)$$

Note that Eq. (7) has been widely adopted to interpret transport in amorphous semiconductors, which at least confirmed that the apparent form of Eq. (7) is correct. The resemblance of Eq. (31) to Eq. (7) simply in terms of their forms is thus an initial validation of the derivations so far. Nevertheless, similarity in the forms does not directly signify that Eq. (31) and Eq. (7) are physically comparable. As a convincing proof, it is required that the properties of the relevant physical quantities should be comparable as well. The fractions in Eq. (31) and Eq. (7) are comparable quantities as they are both determined by the DOS distribution and Fermi statistics; the only difference between the two lies in the energy position, from which states start to be treated as being extended or localized. The question now is whether the artificially defined quantity $\mu_{Ceff}$ in Eq. (31) and the extended-state mobility $\mu_C$ in Eq. (7) are comparable.

A semiquantitative attempt is made on this matter by simply looking into the temperature dependence of these two quantities. $\mu_C$ in the abrupt mobility edge model quantifies the mobility of an extended-state electron at the mobility edge. Similar to the electron mobility in crystalline materials that is influenced by phonon scattering, $\mu_C$ is believed to vary contrarily with temperature; Mott theoretically derived the relation to be $\mu_C \propto T^{-1}$ [27]. It is thus interesting to investigate if the artificially defined $\mu_{Ceff}$ would behave similarly. A key fact to be noted is that $n_{ext,eff}$, $n_{loc,eff}$ and $n_{loc,eff}'$ all increase with an increasing temperature, but their rates of increase are different. It is perhaps easier to first compare the temperature dependence of $n_{loc,eff}$ and $n_{loc,eff}'$. According to the inner product in Eq. (29) and considering the fourth attribute of $\alpha$, $n_{loc,eff}'$ increases more rapidly than $n_{loc,eff}$ as temperature increases. Next, to compare the temperature dependence of $n_{ext,eff}$ and $n_{loc,eff}$, a rough quantification of them is made based on an assumption of a single conducting level and a single trapping level which was widely adopted in the last century [13,14,21]. As will be pointed out in Sec. III, this assumption is not accurate but it suffices for the semiquantitative comparison here, which leads to the approximations that

$$n_{ext,eff} \approx g(E_{Ceff})f(E_{Ceff})kT, \quad (32)$$

and

$$n_{loc,eff} \approx g(E_T)f(E_T)kT, \quad (33)$$

where $E_T$ is the equivalent single trapping level in the band tail. Understanding Eq. (32) is straightforward given the fact that the Fermi level in an undoped amorphous semiconductor is far below $E_{Ceff}$ such that the majority of electron carriers above $E_{Ceff}$ condense near $E_{Ceff}$. As a rough estimation, Eq. (33) neglects localized electrons in deep defect and impurity states, and a further approximation is made based on the DOS distribution feature exemplified by the $a$-Si:H sample in the Fig. 3 of Ref. [28], where a turn-over region exists in the band tail, leading to a peak at $E_T$ in the energy distribution of electron density (see Fig. 1 of

Ref. [14]). As a result, the ratio $n_{loc,eff}/n_{ext,eff}$ is approximated by

$$\frac{n_{loc,eff}}{n_{ext,eff}} \approx \frac{g(E_T)}{g(E_{Ceff})}\exp\left(\frac{E_{Ceff} - E_T}{kT}\right), \quad (34)$$

where the Fermi-Dirac distribution has been approximated by a Boltzmann distribution. Clearly, $n_{ext,eff}$ is more heavily dependent on temperature than $n_{loc,eff}$ such that the ratio $n_{loc,eff}/n_{ext,eff}$ decreases with an increasing temperature in an exponential manner. Given that $n_{loc,eff}'$ increases with temperature more rapidly than $n_{loc,eff}$, the temperature dependence of $n_{loc,eff}'/n_{ext,eff}$ is therefore weaker than $n_{loc,eff}/n_{ext,eff}$. Based on a reasonable assumption that both $\mu_0$ and $\mu_{Cl}$ are also only weakly dependent on temperature, the temperature dependence of $\mu_{Ceff}$ as a whole should take a form that is weaker than the exponential in Eq. (34). Consequently, it is not unacceptable that the temperature dependence of $\mu_{Ceff}$ in Eq. (31) is comparable to that of $\mu_C$ in Eq. (7) as the latter follows an inverse power-law form which is weaker than the exponential form.

### F. Applicability to Non-Dispersive Conduction in a Time-of-Flight Experiment

So far the derivations have been targeting steady-state DC conduction in amorphous semiconductors. Their applicability to the non-dispersive TOF conduction, which is in a similar steady state, is now discussed. Table I summarizes the main differences between the non-dispersive drift mobility and the DC conductivity mobility. Based on these, the earlier derivations in the context of DC conduction are reconsidered.

The first quantitative consideration is given to the probability terms $P_i$, $P_{ijD}$ and $P_{ijIu}$ in Eqs. (10), (16) and (17). Two facts are noted. First, photogenerated electrons initially exist within a narrow "sheet" in the material due to the narrow absorption depth of the exciting laser [21]. Second, these photogenerated electrons will occupy a portion of available basic electronic states along with the intrinsic electron carriers in the material. However, as the TOF measurement progresses, the sheet of photogenerated electrons spreads out and the separation between the photogenerated electrons increases. As a result, for any photogenerated electron, the number of interactable unoccupied basic electronic states is negligibly affected by the presence of other photogenerated electrons because they are trivial compared with the intrinsic carriers that have already existed in the material. Consequently, in the probability terms, it is still the Fermi-Dirac function $f$, rather than the occupation number $F$, that dominates the distribution of unoccupied basic electronic states. Nevertheless, considering the difference between non-dispersive TOF conduction and DC conduction, two modifications are:

(1) The summation index $i$ now only includes those energy levels that are above the thermalization depth.
(2) The total number of relevant levels below $E_{Ceff}$ are $m'$ which is fewer than $m$.

TABLE I. Main differences between non-dispersive drift mobility and DC conductivity mobility

|  | Non-Dispersive Drift mobility | DC conductivity mobility |
| --- | --- | --- |
| Contribution | Photogenerated electrons above the thermalization depth $E_b$ or $E_D$ | All electron carriers intrinsic to the material |
| Measurement time scale | Transient (typically ~ 100 ns) [20] | Sufficiently long to achieve a steady signal |
| Conduction dynamics | A negligible period of nonequilibrium followed by a predominant period of quasi thermal equilibrium | Complete thermal equilibrium |
| Electron occupation function | $F$: above the thermalization depth, $F$ is a Boltzmann distribution extended from the thermalization depth. | $f$: Fermi-Dirac distribution centered at the dark Fermi level $E_F$. |

The second quantitative consideration is on the analysis of thermal equilibrium in Sec. II C. Now targeting quasi thermal equilibrium and its associated transition balance, the analysis which leads to Eq. (22) still holds true. However, instead of targeting intrinsic carriers, the focus moves to the group of photogenerated electrons within an effective volume of $V_p$ in which these drifting electrons prevail. As a result, the occupation number $F(E_i)$, being a stable function, replaces the first Fermi-Dirac term $f(E_i)$ in Eq. (22). Moreover, the emission rate in the volume $V_p$ near $E_i$ is defined and is denoted as $R_V(E_i)$ which only targets photogenerated electrons and replaces the emission rate $R(E_i)$ in Eqs. (18) and (21) which targets intrinsic carriers in a unit volume. Thus, Eq. (22) is modified to

$$F(E_i)g(E_i)\Delta E = C_V(E_i)\sum_{j=1}^{n}\left[t_r(E_i,E_j)P_{ijD}'\right.$$
$$\left.+\sum_{u=1}^{i-1}\{[t_r(E_i,E_u)+t_r(E_u,E_j)]P_{ijIu}'\}\right]. \quad (35)$$

where $P_{ijD}'$ and $P_{ijIu}'$ denote the modified probability terms, and $C_V(E_i)$ expresses the rate of capturing photogenerated electrons from $E_{Ceff}$ to $E_i$, which is quantified by

$$C_V(E_i) = V_p\left[\sum_{j=1}^{n}F(E_j)g(E_j)\Delta E\right]v_{th}\sigma(E_{Ceff},E_i)[1-f(E_i)]g(E_i)\Delta E. \quad (36)$$

Note that like the intrinsic electrons in DC conduction, any photogenerated electron that is released to an energy higher than $E_{Ceff}$ during the TOF measurement will have already relaxed to $E_{Ceff}$ before being subsequently captured to an energy lower than $E_{Ceff}$. Hence, the capture cross section $\sigma$ in Eq. (36) is also always correlated with $E_{Ceff}$ rather than other energies.

All these three reconsiderations eventually lead to the expression of the non-dispersive drift mobility in a steady TOF experiment; this is

$$\mu_d = \mu_{Ceff}\frac{N_{ext,eff}}{N_{ext,eff}+N_{loc,eff}}, \quad (37)$$

where the number of photogenerated electrons that are in effective extended states and those that are in effective localized states are respectively quantified by

$$\begin{cases}N_{ext,eff} = V_p\sum_{j=1}^{n}[F(E_j)g(E_j)\Delta E]\\N_{loc,eff} = V_p\sum_{i=1}^{m'}[F(E_i)g(E_i)\Delta E]\end{cases}. \quad (38)$$

It is clear that the effective volume $V_P$ only serves the purpose of assisting the microscopic analysis but quantitatively knowing its value is entirely unnecessary as it is canceled in the end.

### G. Key Outcomes

First, it is worthwhile to note that quite a few microscopic physical quantities occurred during the earlier derivations to assist understanding and reveal the detailed carrier dynamics. However, thanks to the establishment of thermal equilibrium or quasi thermal equilibrium, none of the following physical quantities directly determines the final mobility result, and so none of them needs to be quantitatively known.

(1) The electron capture cross section $\sigma(E_{Ceff},E_i)$ which is energy specific.
(2) The emission coefficient $e(E_i,E_j)$ and the effective emission coefficient $e'(E_i,E_j)$ which also depend on energies.
(3) The attempt-to-escape frequency $v$.
(4) The energy-specific thermal release time such as $t_r(E_i,E_j)$.

Second, similar to the analysis at the end of Sec. II E, it can also be confirmed that Eq. (37) is comparable with Eq. (5) for non-dispersive TOF conduction. As will be shortly shown in Sec. III, in previous literature, Eq. (5) served as the basis for the experimental determination of two key parameters: extended-state mobility and mobility edge. Comparing Eq. (37) and Eq. (5), brand-new understanding is yielded.

(1) The experimentally determined extended-state mobility should actually be $\mu_{Ceff}$, which, according to Eq. (30), is higher than the actual free electron mobility $\mu_0$; this is a direct consequence of the fact that $\alpha(E)$ cannot be neglected. This has not been considered before because of the stereotypical model of an abrupt mobility edge below which states are completely confined.
(2) The experimentally determined mobility edge, being the $E_C$ associated with Eq. (6), should actually refer to the effective mobility edge $E_{Ceff}$ associated with Eq. (38); this is different from (and likely to be higher than) the critical level $E_{ct}$ which demarcates the actual extended states and the actual localized states.

Section III will use a typical $a$-Si:H thin film as an example and estimates the values of these two parameters for this material, which concretizes the above key results.

### H. Extending from the Ideal Band Fluctuation Model to the Actual Band Fluctuation Model

Returning to Fig. 2, it is recognized that the highly symmetric band fluctuation model adopted so far does not entirely reflect the actual situation shown in Fig. 2(b). The actual band edge spatial distribution is a result of a semi-random spatial rearrangement of the SRLs in Fig. 2(f); the only principle that limits the spatial rearrangement is the requirement that the fluctuation of band edges is slow, which is determined by the short- and medium-range order in amorphous semiconductors (see the prequel paper [23] for details). However, the band fluctuation scale exhibited in Fig. 2(b) is in fact far shorter than the size of a realistic thin film sample (which is not modeled due to excessive computational loads). Thus, the common feature shared by the ideal model and the actual model is that they are both *macroscopically uniform*; it is only that for the ease of analysis, the ideal model was adopted which further ensures microscopic uniformity that justifies the use of macroscopic quantities (e.g., the DOS) in the microscale analysis. Admittedly, the actual microscopic nonuniformity does complicate the transport analysis. A rigorous quantification requires consideration of not only the detailed local electronic structures along the path of a single electron but also the different electronic structures that different electrons interact with.

However, it is still believed that the microscopically uniform model provides a good approximation given that the material under investigation is assumed to be sufficiently large. The direct consequence of this is that, during conduction, the majority of electrons have opportunities to interact with the basic electronic states at most energy levels such that they end up exhibiting similar mobilities. It may still appear ambiguous to analyze the migration of one of these electrons since the downward transition and release probabilities are spatially varying given the microscopic nonuniformity. Nevertheless, taking the downward transition probability as an example, while it is easier for an electron to transition to a state at locations where this state condenses (corresponding to a higher local DOS than the macroscopic average), it is more difficult or impossible for this electron to transition to this state at locations where this state is trivial or absent (corresponding to a local DOS lower than the macroscopic average). These two scenarios counteract each other such that assuming a uniform transition probability everywhere based on the macroscopic DOS remains an acceptable approximation. This is on condition that the number of MTR cycles is sufficiently large to allow the exhibition of statistical nature; in the setting of a large thin film material, this condition is satisfied. Similar consideration applies to the release probabilities.

The detailed difference from the actual situation may lead to quantitative mismatches between the ideal model and the actual model. For instance, the exact position of $E_{Ceff}$ may differ for the two models. Also, the quantification of the drift efficiency $\alpha(E)$ may be different, because the extent of confinement at the same energy can differ for the two models [see Fig. 5(a) – (c)]. This means that the value of $\mu_{Ceff}$ can be different in the two models. Nevertheless, the key outcome, which is Eq. (31) or (37), remains entirely valid for the actual model due to the macroscopic uniformity. The exact position of $E_{Ceff}$ and the exact value of $\mu_{Ceff}$ are not to be determined in a bottom-up manner based on model-specific parameters; instead, they should be reversely determined based on experimental evidence. The next section concretizes $E_{Ceff}$ and $\mu_{Ceff}$ using an $a$-Si:H sample as an example.

### III. DETERMINATION OF EFFECTIVE MOBILITY EDGE AND EFFECTIVE EXTENDED-STATE MOBILITY OF AN $a$-Si:H SAMPLE BASED ON EXISTING EXPERIMENTAL DATA

To ensure coherence, focuses in this section are given to a historically well-studied $a$-Si:H sample; there are associated DOS data [28] and temperature-dependent TOF data [20] that were characterized by the same reputable laboratory around the same time. This is also to ensure coherence with the prequel paper [23], which is mainly based on the same DOS data. This DOS data is preferred because it was obtained using field-effect technique which is good at revealing DOS in the important band tail region and which cannot be superseded by even the later wide-range DOS measurement techniques (e.g., Ref. [29]).

#### A. Effective Mobility Edge $E_{Ceff}$

According to Eq. (31), the DC conductivity $\sigma_{dc}$ is quantified by $e\mu_{Ceff}n_{ext,eff}$, where $e$ is the elementary charge. To roughly reveal the temperature dependence of $\sigma_{dc}$, the single conducting level approximation in Eq. (32) is adopted which yields:

$$\sigma_{dc} \approx e\mu_{Ceff} g(E_{Ceff})kT \exp\left[-\frac{(E_{Ceff} - E_F)}{kT}\right]. \quad (39)$$

Here, considering the statistical shift, the Fermi level may be quantified as $(E_{F0} + \gamma_F T)$, where $E_{F0}$ denotes the Fermi level at 0K and $\gamma_F$ is a factor quantifying the dependence on temperature $T$ [18,30]. It has been analyzed that the temperature dependence of $\mu_{Ceff}$ is weak. Therefore, in Eq. (39), the temperature dependence of conductivity $\sigma_{dc}$ is dominated by $\exp[-(E_{Ceff} - E_{F0})/kT]$. As a result, the relative position of $E_{Ceff}$ can be easily deduced through a measurement of activation energy $(E_{Ceff} - E_{F0})$.

The value of activation energy is important in the characterization of DOS distribution using field effect technique, because it specifies the relative energy position of the DOS curve [31]. However, previous beliefs were that the activation energy is associated with the critical energy $E_{ct}$ which distinguishes extended states and localized states [31]. As a result, the energy reference of the measured DOS distribution was assumed to be at $E_{ct}$ as well. Now with the new understanding in this paper, it is clear that the energy reference should actually be at $E_{Ceff}$. Thus, for the $a$-Si:H sample in Ref. [28] (which is also the same sample we have modeled in the prequel paper [23]), $E_{Ceff}$, being the energy reference, is 0.14 eV higher than the upper boundary of the exponential DOS region; this, in the model of the prequel [23], means that $E_{Ceff}$ is 0.074 eV higher than $E_{ct}$. Back to the start of Sec. II, it has been qualitatively envisaged based on Fig. 2(c) that $E_{Ceff}$ is likely to be higher than $E_{ct}$ for the studied $a$-Si:H; this is now retrospectively validated.

#### B. Effective Extended-State Electron Mobility $\mu_{Ceff}$

In Sec. II, the assumption of a single conducting level and single trapping level allows a rough estimation of the density of electron carriers that are in effective extended states and effective localized states, respectively, in the context of steady DC conduction. When used in non-dispersive TOF conduction, this assumption straightforwardly yields that the drift mobility exhibits a temperature dependence that is [13,14]

$$\mu_d = \mu_{Ceff} \frac{1}{1 + \frac{g(E_T)}{g(E_{Ceff})}\exp\left(\frac{E_{Ceff} - E_T}{kT}\right)}. \quad (40)$$

Fitting Eq. (40) to experimental temperature-dependent drift mobility data, the values of $\mu_{Ceff}$, $E_T$ and the DOS ratio $g(E_T)/g(E_{Ceff})$ can be easily obtained. The typical values of $\mu_{Ceff}$ fitted this way are within the range from 10 to 20 cm$^2$/(V s) for the historical $a$-Si:H sample [13,14,18].

The adoption of the single-level approximation simplified the analysis and meanwhile provided a very simple analytical form of drift mobility which could be easily used to fit to the experimental temperature-dependent data. However, underneath this convenience, the nature of the continuity in the DOS distribution was overlooked. To accurately reflect the actual situation, this paper abandons the single level approximation, and revisits the fitting to TOF data from the viewpoint of the actual continuously distributed DOS. The same $a$-Si:H sample is focused whose DOS in the important band tail region were reliably measured through field-effect technique [28]. To span a wider energy range towards the vicinity of $E_{Ceff}$ and above, experimental evidence from other techniques is incorporated which eventually qualifies the DOS result in the prequel to this paper

[23] (copied in Fig. 6(a) with $E_C$ revised to $E_{Ceff}$) as an excellent representation of the actual situation. This DOS distribution is segmented into four parts, which are respectively expressed as:

$$g(E) = g(E_0) \exp\left(\frac{E - E_0}{kT_C}\right), E_D < E < E_0, \quad (41)$$

$$g(E) = g(E_0) \exp\left(\frac{E - E_0}{kT_0}\right), E_0 < E < E_1, \quad (42)$$

$$g(E) = hE + w, E_1 < E < E_2, \quad (43)$$

or,

$$g(E) = A_1\sqrt{E - A_2}, E > E_2, \quad (44)$$

Equations (41), (43) and (44) respectively describe the exponential region, linear region and parabolic region labeled in Fig. 6(a), respectively, while Eq. (42), for the propose of computational simplicity, uses a second exponential function to satisfactorily approximate the transitional region between the exponential and linear region. The characteristic slopes of the two exponential regions are $T_C$ and $T_0$, respectively. These parameters, together with the linear function parameters $h$ and $w$ as well as the parabolic function parameters $A_1$ and $A_2$, can all be directly extracted from Fig. 6(a). Note that these structural parameters are fixed constants that faithfully reflect the actual situation, which are listed in Appendix B.

A reasonable view to explain the fact that this $a$-Si:H sample exhibits non-dispersive features even down to ~ 150 K [14,20] is that, as discussed in Sec. I, the capture cross section drops abruptly below $E_D$, which was also corroborated in later studies (e.g., Ref. [15]). Here in this paper, an assumption is that, $E_D$, which defines the thermalization depth, is lower than the energy ($E_0$) where the two exponential regions meet. This will be retrospectively validated by the subsequent fitting. The lower limit of $E_D$ is assumed to be at ~ 0.23 eV below $E_{Ceff}$. This assumption is to ensure that Eq. (41) is valid throughout its energy range; otherwise, if $E_D$ is noticeably lower, deeper DOS needs to be separately quantified, because, as revealed by the actual field-effect data, charged defect states slow down the DOS variation [28,32]. This assumption will also be retrospectively validated.

It is analyzed in Sec. II that the temperature dependence of $\mu_{Ceff}$ is comparable to that of $\mu_C$ in the abrupt mobility edge model; the latter, as mentioned earlier, was theoretically derived by Mott as being inversely proportional to temperature [27]. Given the possible inaccuracies brought by the embedded approximations as well as by the previous understanding of localized states, it is more generally assumed in this paper (as well as in the prequel paper) that $\mu_{Ceff} \propto T^{-p}$ where $p \sim 1$ but is not necessarily equal to unity. Hence,

$$\mu_{Ceff} = \mu_{CR}\left(\frac{T_R}{T}\right)^p, \quad (45)$$

where $\mu_{CR}$ denotes the effective extended-state electron mobility at room temperature $T_R = 300$ K.

Combining Eqs. (41) to (44) (i.e., the continuous DOS information) with Eqs. (37), (38) and (45) (i.e., the modified MTR model), fitting to the TOF data on exactly the same $a$-Si:H sample in Ref. [20] (or equivalently, Ref. [14]) is conducted. Detailed analytical and numerical calculations are given in Appendix B. The result is shown here in Fig. 6(b). The fitting is excellent over the wide temperature range from 167 K to 457 K. Fitting parameters are determined which are discussed below.

The room-temperature effective extended-state electron mobility $\mu_{CR}$ is fitted to be 46 cm$^2$/(V s), in contrast to the lower values based on the single level approximation. The most noticeable issue of adopting the rough single level approximation lies in the quantification of electron density using an energy spread of $kT$. In comparison, integration over the actual continuously distributed DOS yields a more reliable result. This result may appear counterintuitive, because the mobility of free electrons slightly above the band tail in $a$-Si:H was believed to just transition from the regime of Brownian motion to the regime of propagation, corresponding to a mobility value of ~ 5 cm$^2$/(V s) [35]. This counter-intuitiveness, however, exactly matches the analysis in Sec. II, which states that the effective extended-state mobility is higher than the actual mobility of free electrons in the amorphous semiconductor.

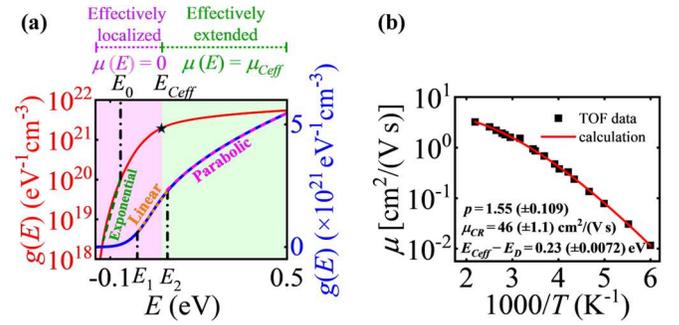

FIG. 6. Quantification of electronic structure and fitting of electronic transport. (a) The DOS distribution calculated in the prequel [23] which accurately reflects the actual situation. The solid red curve and blue curve exhibit the DOS distribution in the logarithmic and linear scale, respectively. The exponential, linear and parabolic regions are labeled with their respective boundaries indicated by $E_0$, $E_1$ and $E_2$. The dashed olive line represents the fitting of the calculation to the field-effect measurement data in Ref. [28]. The star symbol denotes the data in Ref. [33] which agrees with inverse photoemission results [34]. Based on Eqs. (31) or (37), it is equivalent to treat electrons above the effective mobility edge $E_{Ceff}$ as in extended states exhibiting a mobility of $\mu_{Ceff}$ and treat electrons below $E_{Ceff}$ as localized with zero mobility. (b) Fitting of the drift mobility model in Eq. (37) to the TOF data in Ref. [20] (equivalently, Ref. [14]). The fitting parameters are determined with their confidence intervals indicated.

The fitted $p$ of 1.55 is justifiable. As a first-order approximation, electron mobility in undoped crystalline semiconductors is inversely proportional to $T^{1.5}$ due to the influence of phonon scattering [36]. Although silicon atoms in $a$-Si:H no longer sit in a periodic lattice, their thermal vibrations are similar to those in crystalline silicon ($c$-Si). Since the average spacing between neighboring silicon atoms in $a$-Si:H and crystalline silicon are also similar [37], the temperature-dependence of the vibrational scattering of free electrons in $a$-Si:H should be similar to that in $c$-Si. Despite the fact that electrons are meanwhile scattered by the fluctuating band due to structural disorder, the extent of disorder (and thus this scattering mechanism) is not as sensitive to temperature as the vibrational scattering. As a result, the overall temperature dependence, quantified by a $p$ close to 1.5, is acceptable for undoped $a$-Si:H.

The thermalization depth of 0.23 eV is obtained, which retrospectively validates the earlier assumptions on the lower and upper limits of $E_D$. According to Eq. (2), this thermalization depth corresponds to a room-temperature thermalization time of 0.87 ns under the assumption that $v = 10^{13}$ s$^{-1}$; this is much shorter than the corresponding total transit time of ~ 100 ns at room temperature [13,28], and is thus able to account for the observed non-dispersive features at room temperature. For other temperatures, there is a balance between the change of total transit time and the change of thermalization time; they are both temperature-dependent, and their relative difference determines whether a transport is non-dispersive or dispersive. According to Eq. (2) and the TOF data in Fig. 6(b), the general trend is that, under the same electric field, the increase of thermalization time is more drastic than the increase of total transit time with a decreasing temperature. The result is that the unstable thermalization remains insignificant relative to the total transit time above ~ 150 K, but could become more noticeable after the temperature further decreases to below ~ 150 K; this is when non-dispersive transport could transition to dispersive transport.

## IV. CONCLUSION

In conclusion, this paper revisits the MTR transport theory of amorphous semiconductors exemplified by $a$-Si:H. The problems of the conventional abrupt mobility edge model are pointed out, and the actual attributes of extended states and localized band tail states in amorphous semiconductors are considered. A more accurate analytical model associated with MTR is developed. Importantly, it is revealed that the experimentally derived mobility edge of an amorphous semiconductor is an effective quantity which is different from the actual critical energy that demarcates extended states and localized states. Further, the experimentally derived extended-state electron mobility is also an effective quantity which turns out to be higher than the actual free electron mobility in the amorphous semiconductor. Next, the authors quantify these two effective quantities using a historically well-studied $a$-Si:H sample as an example. It is found that for this $a$-Si:H, the effective mobility edge is estimated to be 0.074 eV higher than the critical energy, and the effective extended-state electron mobility at room temperature is 46 cm$^2$/(V s) that is higher than the free electron mobility of ~ 5 cm$^2$/(V s). In the meantime, the authors point out that the single level approximation that was widely adopted in the past is not accurate for the purpose of determining the effective extended-state mobility. Instead, the actual continuously distributed DOS should be used. The MTR is an important mechanism dominating the transport of a variety of amorphous semiconductors that are of high technological importance, and so the insights in this work are significant in optimizing existing devices and proposing new devices based on amorphous semiconductors. The sequel to this paper [24] will build on this paper and its prequel, and extend the MTR theory into the nanoscale.

## ACKNOWLEDGEMENT

This work is supported by the UKRI Engineering and Physical Sciences Research Council under grant number EP/W009757/1. The authors acknowledge the Rank Prize for their Return to Research Grant. Y.L. is grateful to the Cambridge Commonwealth, European and International Trust for their Ph.D. scholarship.

## APPENDIX A: CONSTRUCTION OF THE IDEAL BAND EDGE FLUCTUATION MODEL

The reference energy in Fig. 2 is defined by the $c$-Si band edge position which is set as 0 eV. Different from the actual band edge fluctuation model, in the ideal model, the deviations of the local band edges from the reference energy are envisaged to be highly ordered for the purpose of providing the analytical simplicity required in Sec. II. The definitions of order are that:

(1) The deviations are highly ordered in space. Regions of positive deviation and regions of negative deviation are alternately and uniformly distributed in space.
(2) The deviations are highly ordered in energy. The maximum local band edge in all positive deviation regions are identical, and the minimum local band edge in all negative deviation regions are also identical.
(3) The variation of local band edge with spatial location is identical for all positive deviation regions. The same holds true for all negative deviation regions.

Another important assumption which further simplifies the construction of the ideal model is that the variation of local band edge within any deviation region is monotonic. It is monotonically decreasing away from the center of any positive deviation region or towards the center of any negative deviation region. This is different from the actual complex model in Fig. 2(a) where there are high-frequency fluctuations between a minimum and a maximum.

In these senses, it is useful to introduce the methodology in crystallography and treat the spatial arrangement of positive deviation regions (henceforth labeled as "+") and negative deviation regions (henceforth labeled as "−") as an analogy to the arrangement of atoms in a Bravais lattice. As a result, a unit cell can be defined in a presumably infinite solid, which contains the fewest "+" and "−" regions and which repeats throughout the three-dimensional (3D) space. Accordingly, the probability density function (PDF) that describes the energetic distribution of local band edges in the entire solid can be used to quantify the energetic distribution of local band edges in any unit cell.

It is necessary to ensure that there must be a definite value of local band edge at every location in the material. Therefore, the basic consideration now is on a strategy which ensures that

(1) The "+" and "−" regions *completely* fill up the material space with no residuals. This is different from the case of crystallography that allows residual space between atoms.
(2) The "+" and "−" regions do not overlap with each other; otherwise there can be two values of local band edge at some location.

The easiest way is to envisage that the "+" and "−" regions are cubes with a side length of $A$ which alternately sits on the 8 vertices of individual unit cells in a simple cubic lattice whose lattice constant is exactly $A$. This is schematically shown in Fig. 7(a). In this case, the boundaries of "+" and "−" regions merge, so the value of local band edge on these boundaries must be identical. In another word, the decrease of local band edge from the center of a "+" region to the center of an adjacent "−" region must be strictly *continuous*. The value of local band edge on these boundaries is henceforth labeled as $E_{CC}$ (not to be confused with the reference energy $E_{C0}$). Given that the boundary of any single "+" or "−" region is a cube surface and that the local band edge varies monotonically with location, it can be envisaged that an iso-energy surface, which specifies the locations that possess a certain value of local band edge, must also be a cube surface. This is illustrated in Fig. 7(a). Accordingly, each of such iso-energy surfaces can be specified by the shortest distance from the center of the corresponding "+" or "−" region; this is denoted as $r$.

Consider the local band edge values that are within a small energy interval $[E_{CL}, E_{CL}+\delta E_{CL}]$ with $E_{CL} > E_{CC}$. The probability of finding short-range localities (SRLs) where the local band edge is within this energy interval is calculated from the PDF as $\rho(E_{CL})\delta E_{CL}$, with $\rho(E_{CL})$ being the PDF value at $E_{CL}$. Therefore, within a unit cell of a volume of $A^3$, the number of SRLs where the local band edge is within $[E_{CL}, E_{CL}+\delta E_{CL}]$ is

$$N_S(E_{CL}) = \rho(E_{CL})\delta E_{CL}\left(\frac{A^3}{a_0^3}\right), \quad (A1)$$

where, as defined in this paper and its prequel, the side length of a SRL is $a_0$. As each "+" region in the unit cell is meanwhile shared by 7 adjacent unit cells, every unit cell contains 0.5 complete "+" region. Consequently, there are $2 \times N_S(E_{CL})$ SRLs inside a *complete* "+" region where the local band edge is within $[E_{CL}, E_{CL}+\delta E_{CL}]$.

These SRLs, according to the monotonicity assumption, exist within a thin shell sandwiched between the iso-energy surface of $E_{CL}$ at coordinate $r$ and the iso-energy surface of $(E_{CL}+\delta E_{CL})$ at coordinate $(r+\delta r)$. This is illustrated in the inset of Fig. 7(a). The number of SRLs included in this thin shell is

$$-\frac{6\times(4r^2)\delta r}{a_0^3}, \quad (A2)$$

where the minus sign stems from the fact that the local band edge decreases away from the center of the "+" region. (A2), based on the monotonicity assumption, is equal to $2\times N_S(E_{CL})$. Therefore,

$$2\times\rho(E_{CL})\delta E_{CL}\left(\frac{A^3}{a_0^3}\right) = -\frac{6\times(4r^2)\delta r}{a_0^3}, \quad (A3)$$

Integrating from the center of the "+" region to its boundary, the value of local band edge varies monotonically from the maximum ($E_{max}$) to $E_{CC}$; this gives

$$\int_{E_{max}}^{E_{CL}(r)} 2\times\rho(E_{CL})\left(\frac{A^3}{a_0^3}\right)dE_{CL} = \int_0^r -\frac{6\times(4r^2)}{a_0^3}dr, \quad (A4)$$

where the boundary condition is that $E_{CL}(A/2) = E_{CC}$.

Similarly, considering the case of $E_{CL} < E_{CC}$, the variation of local band edge in a "−" region can be derived as

$$\int_{E_{min}}^{E_{CL}(r)} 2\times\rho(E_{CL})\left(\frac{A^3}{a_0^3}\right)dE_{CL} = \int_0^r \frac{6\times(4r^2)}{a_0^3}dr, \quad (A5)$$

where $E_{min}$ denotes the minimum local band edge in the "−" region and $r$ indexes the location relative to the center of the "−" region which ranges from 0 to $A/2$. The boundary condition is still that $E_{CL}(A/2) = E_{CC}$. Combining the boundary conditions of Eqs. (A4) and (A5), it is easy to find that

$$\int_{E_{CC}}^{E_{max}}\rho(E_{CL})dE_{CL} = \int_{E_{min}}^{E_{CC}}\rho(E_{CL})dE_{CL} = \frac{1}{2}. \quad (A6)$$

Therefore, $E_{CC}$ is the median of the PDF, which equally demarcates "+" and "−" regions both spatially and in terms of the energetic distribution.

Figure 7(b) is a result of numerical integration based on the known PDF in the prequel paper [23] (adopted in this paper as well). It shows the one-dimensional (1D) variation of local band edge between adjacent "−" and "+" regions (assuming that the coordinate $r_x$ moves from the center of a "−" region straight towards the center of a nearest "+" region). Based on this 1D variation, Fig. 2(e) is obtained which shows the two-dimensional (2D) band edge variation in a plane that contains the top surfaces of unit cells.

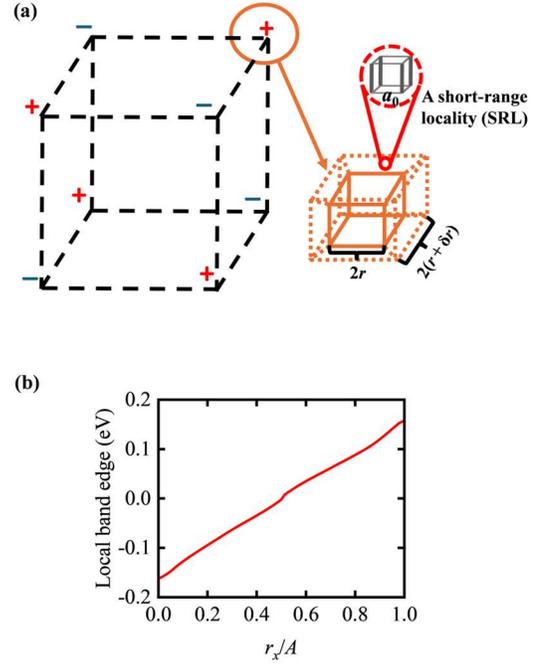

FIG. 7. Method of constructing the ideal band edge fluctuation model. (a) Ideal spatial arrangement of positive deviation regions ("+") and negative deviation regions ("−"). The centers of these individual regions sit within the framework of a simple cubic lattice. The value of local band edge decays away from the centers of individual "+" regions towards the centers of individual "−" regions. The iso-energy surfaces are assumed to be cube surfaces. The coordinate ($r$) of an iso-energy surface is defined as the shortest distance from the nearest deviation center. A thin shell is sandwiched between the iso-energy surface at $r$ and the iso-energy surface at $r+\delta r$, within which a certain number of short-range localities (SRLs) are spatially included. (b) Calculated one-dimensional (1D) variation of local band edge between adjacent "−" and "+" regions. $r_x$ denotes the distance away from the center of a "−" region. This is a numerical integration result based on the known probability density function (PDF) in Fig. 2(d). Expansion of this result to a two-dimensional (2D) plane leads to the results in Fig. 2(e) and (f).

Though being highly ideal, this model is macroscopically comparable to the actual model. As discussed in Sec. II H, the actual model is merely treated as a result of stochastic rearrangement of SRLs, where certain SRLs may congregate together leading to the non-uniform condensation of some localized band tail states. Such microscopic differences, however, do not violate the key results derived in Sec. II based on the ideal model.

## APPENDIX B: DETAILS OF FITTING USING CONTINUOUSLY DISTRIBUTED DOS

According to Eqs. (41) to (44) and Eqs. (37), (38), and with the knowledge that photogenerated electrons rapidly reaches an equilibrium above $E_D$, the non-dispersive electron drift mobility is expressed as

$$\mu_d = \mu_{Ceff} \frac{N_1 + N_2}{N_1 + N_3 + N_4 + N_5}, \quad (B1)$$

Here, $N_1$ and $N_2$ evaluate the number of photogenerated electrons that are in effective extended states; they respectively represent the parabolic region above $E_2$ and the energy range between $E_{Ceff}$ and $E_2$ (in the linear region). The total number of photogenerated electrons that contribute to the TOF conduction is evaluated by $N_1 + N_3 + N_4 + N_5$ with $N_3$, $N_4$ and $N_5$ respectively representing the entire linear region, the second and the first exponential region. After preliminary mathematical reductions,

$$N_1 = V_P A_1 (kT)^{\frac{3}{2}} \exp\left(\frac{E_D - A_2}{kT}\right) \Gamma\left(\frac{3}{2}\right) \left[1 - \Gamma_{inc}\left(\frac{3}{2}, \frac{E_2 - A_2}{kT}\right)\right], \quad (B2)$$

$$N_2 = V_P h (kT)^2 \exp\left(\frac{E_D}{kT}\right) \Gamma(2) \left[\Gamma_{inc}\left(2, \frac{E_2}{kT}\right) - \Gamma_{inc}\left(2, \frac{E_{Ceff}}{kT}\right)\right] + V_P \exp\left(\frac{E_D}{kT}\right) wkT \left[\exp\left(-\frac{E_{Cef}}{kT}\right) - \exp\left(-\frac{E_2}{kT}\right)\right], \quad (B3)$$

$$N_3 = V_P h (kT)^2 \exp\left(\frac{E_D - E_1}{kT}\right) \left\{\Gamma(2)\Gamma_{inc}\left(2, \frac{E_2 - E_1}{kT}\right) + hE_1 kT \left[1 - \exp\left(\frac{E_1 - E_2}{kT}\right)\right]\right\} + V_P \exp\left(\frac{E_D}{kT}\right) wkT \left[\exp\left(-\frac{E_1}{kT}\right) - \exp\left(-\frac{E_2}{kT}\right)\right], \quad (B4)$$

$$N_4 = V_P g(E_0) \exp\left(\frac{E_D}{kT} - \frac{E_0}{kT_0}\right) \frac{kTT_0}{T - T_0} \left[\exp\left(\frac{E_1}{kT_0} - \frac{E_1}{kT}\right) - \exp\left(\frac{E_0}{kT_0} - \frac{E_0}{kT}\right)\right], \quad (B5)$$

$$N_5 = V_P g(E_0) \exp\left(\frac{E_D}{kT} - \frac{E_D}{kT_C}\right) \frac{kTT_C}{T - T_C} \left[\exp\left(\frac{E_0}{kT_C} - \frac{E_0}{kT}\right) - \exp\left(\frac{E_D}{kT_C} - \frac{E_D}{kT}\right)\right], \quad (B6)$$

Here, $\Gamma_{inc}$ in Eqs. (B2), (B3) and (B4) denotes the lower incomplete gamma function of the form

$$\Gamma_{inc}(s_0, x_0) = \frac{1}{\Gamma(s_0)} \int_0^{x_0} z^{s_0 - 1} \exp(-z) \, dz, \quad (B7)$$

with $\Gamma(s_0)$ being

$$\Gamma(s_0) = \int_0^{+\infty} z^{s_0 - 1} \exp(-z) \, dz. \quad (B8)$$

In these two equations, $z$ is an independent variable. Fixed constants extracted from the DOS distribution are: $E_0 = -0.065$ eV, $E_1 = -0.008$ eV, $E_2 = 0.094$ eV, $E_{Ceff} = 0.075$ eV, $A_1 = 7.735 \times 10^{21}$ eV$^{-3/2}$cm$^{-3}$, $A_2 = 0.00307$ eV, $h = 1.65327 \times 10^{22}$ eV$^{-2}$cm$^{-3}$, $w = 7.23683 \times 10^{20}$ eV$^{-1}$cm$^{-3}$, $g(E_0) = 10^{20}$ eV$^{-1}$cm$^{-3}$, $T_0 = 371.468$ K, and $T_C = 201$ K. Fitting is done in OriginPro® 2021.